\documentclass{emulateapj} 

\gdef\rfUV{(U-V)_{\rm rest}}
\gdef\rfVJ{(V-J)_{\rm rest}}
\gdef\1054{MS\,1054--03}

\def\farcs{\hbox{$.\!\!^{\prime\prime}$}}
\def\simgeq{{\raise.0ex\hbox{$\mathchar"013E$}\mkern-14mu\lower1.2ex\hbox{$\mathchar"0218$}}}

\begin {document}

\title {FIREWORKS $U_{38}$-to-24 $\mu$\lowercase{m} photometry of the GOODS-CDFS: multi-wavelength catalog and total IR properties of distant $K_s$-selected galaxies}

\author{Stijn Wuyts\altaffilmark{1,2}, Ivo Labb\'{e}\altaffilmark{3}, Natascha M. F\"{o}rster Schreiber\altaffilmark{4}, Marijn Franx\altaffilmark{2}, Gregory Rudnick\altaffilmark{5}, Gabriel B. Brammer\altaffilmark{6}, Pieter G. van Dokkum\altaffilmark{6}}
\altaffiltext{1}{Harvard-Smithsonian Center for Astrophysics, 60 Garden Street, Cambridge, MA 02138; W. M. Keck Postdoctoral Fellow [e-mail: swuyts@cfa.harvard.edu]}
\altaffiltext{2}{Leiden Observatory, P.O. Box 9513, NL-2300 RA, Leiden, The Netherlands.}
\altaffiltext{3}{Carnegie Observatories, 813 Santa Barbara Street, Pasadena, CA 91101; Hubble Fellow}
\altaffiltext{4}{MPE, Giessenbackstrasse, D-85748, Garching, Germany}
\altaffiltext{5}{National Optical Astronomical Observatory, 950 North Cherry Avenue, Tucson, AZ 85721; Leo Goldberg Fellow}
\altaffiltext{6}{Department of Astronomy, Yale University, New Haven, CT 06520-8101}

\begin{abstract}
We present a $K_s$-selected catalog, dubbed FIREWORKS, for the Chandra
Deep Field South (CDFS) containing photometry in $U_{38}$, $B_{435}$,
$B$, $V$, $V_{606}$, $R$, $i_{775}$, $I$, $z_{850}$, $J$,
$H$, $K_s$, $[3.6\mu$m$]$, $[4.5\mu$m$]$, $[5.8\mu$m$]$,
$[8.0\mu$m$]$, and the MIPS $[24\mu$m$]$ band.  The imaging has a
typical $K_{s,{\rm AB}}^{\rm tot}$ limit of 24.3 mag ($5\sigma$) and
coverage over 113 arcmin$^2$ in all bands and 138 arcmin$^2$ in all
bands but $H$.  We cross-correlate our catalog with the 1 Ms X-ray
catalog by Giacconi et al. (2002) and with all available spectroscopic
redshifts to date.  We find and explain systematic differences in a
comparison with the '$z_{850}$ + $K_s$'-selected GOODS-MUSIC
catalog that covers $\sim$90\% of the field.  We exploit the
$U_{38}$-to-24 $\mu$m photometry to determine which $K_s$-selected
galaxies at $1.5<z<2.5$ have the brightest total IR luminosities and
which galaxies contribute most to the integrated total IR emission.
The answer to both questions is that red galaxies are dominating in
the IR.  This is true no matter whether color is defined in the
rest-frame UV, optical, or optical-to-NIR.  We do find however that
among the reddest galaxies in the rest-frame optical, there is a
population of sources with only little mid-IR emission, suggesting a
quiescent nature.
\end{abstract}


\section {Introduction}
\label{intro.sec}
Since the original Hubble Deep Field (Williams et al. 1996), deep
multi-wavelength observations of blank fields have revolutionized our
understanding of the high-redshift universe.  Especially the epoch
around $z \sim 2$ is of great interest since it is then that the
cosmic star formation rate density was peaking (Hopkins \& Beacom
2006).  At $z \sim 2$, the observed optical probes the redshifted
rest-frame UV emission, making it a good tracer for relatively
unobscured star formation.  Near-Infrared (NIR) observations of
distant galaxies, such as undertaken by the FIRES survey in the Hubble
Deep Field South (HDFS, Labb\'{e} et al. 2003, hereafter L03) and the
\1054 field (F\"{o}rster Schreiber et al. 2006, hereafter FS06), show
relatively small variations in the rest-frame $V$-band mass-to-light
ratio.  Selecting galaxies in the $K_s$-band (e.g., L03; FS06) thus
provides a good probe of the massive galaxy content at high redshift.

In the presence of dust, large amounts of rest-frame UV emission can
be absorbed and re-emitted in the Far-Infrared (FIR).  Dust
corrections of the UV luminosities of such systems involve large
uncertainties.  Direct observations of the dust emission are therefore
crucial to get a better census of the bolometric energy output.  In
order to study the bolometric properties of typical galaxies at $z
\sim 2$, infrared luminosities have been derived from the observed 24
$\mu$m flux by means of IR spectral energy distribution (SED)
templates (e.g., Papovich et al. 2005; Reddy et al. 2006).

Despite the extra model uncertainty involved, this approach adds
complementary information to the shorter wavelength studies of
high-redshift galaxies.  While the Luminous ($10^{11} L_{\sun}
<L_{\rm IR}<10^{12} L_{\sun}$) and Ultraluminous ($L_{\rm IR}>10^{12}
L_{\sun}$) Infrared Galaxies, (U)LIRGS (Sanders \& Mirabel 1996) are
locally very rare, they are found to be increasingly more common
toward higher redshifts (e.g., Caputi et al. 2006).

In this paper, we present a $K_s$-band selected multi-wavelength
catalog for the GOODS-CDFS, comprising WFI $U_{38}BVRI$, ACS
$B_{435}V_{606}i_{775}z_{850}$, ISAAC $JHK_s$, IRAC
3.6-8.0 $\mu$m and MIPS 24 $\mu$m imaging.  We adopt a similar format
as for the publicly available FIRES catalogs of the HDFS and \1054,
hence the name FIREWORKS.  This allows the user to exploit the
combined photometry of the CDFS,
\1054, and the HDFS in a straightforward manner.  The fields are
complementary in depth ($5\sigma$ for point sources $K_{s,{\rm
AB}}^{\rm tot}=24.3$, $K_{s,{\rm AB}}^{\rm tot}=25.0$, $K_{s,{\rm
AB}}^{\rm tot}=25.6$ respetively) and area (138 arcmin$^2$, 24
arcmin$^2$, and 5 arcmin$^2$ respectively).

An analysis of the space density and colors of massive galaxies at
$2<z<3$ (van Dokkum et al. 2006), the rest-frame optical luminosity
density and stellar mass density up to $z \sim 3$ (Rudnick et
al. 2006), and the rest-frame luminosity functions of galaxies at
$2<z<3.5$ (Marchesini et al. 2006) were partly based on the FIREWORKS
catalog for the GOODS-CDFS presented here.

After describing the catalog construction, we particularly address the
questions which $K_s$-selected galaxies at $1.5<z<2.5$ are brightest
at 24 $\mu$m, which galaxies have the largest total infrared
luminosity $L_{\rm IR}$ ($\equiv L(8-1000\ \mu$m$)$) and contribute most to
the total integrated IR luminosity emitted by $K_s$-selected galaxies.
We address this question by studying the IR emission as function of
color defined in three wavelength regimes: the rest-frame UV, optical,
and optical-to-NIR.

An overview of the observations is presented in
\S\ref{observations.sec}.  \S\ref{finalim.sec} describes the
construction of the final mosaics.  Source detection and photometry is
discussed in \S\ref{det_phot.sec}.  Next, we present our photometric
redshifts ($z_{\rm phot}$) and cross-correlation with the available
spectroscopic surveys in \S\ref{redshifts.sec}.  \S\ref{param.sec}
summarizes the catalog content.  A photometric comparison for the
wavelength bands in common with the GOODS-MUSIC catalog by Grazian et
al. (2006a, hereafter G06a) and a $z_{\rm phot}$ comparison with the
same authors is discussed in \S\ref{comparison.sec}.  Results on 24
$\mu$m properties and total infrared luminosities of $K_s$-selected
galaxies at $1.5<z<2.5$ are discussed in \S\ref{science.sec}.
\S\ref{summary.sec} summarizes the paper.

AB magnitudes are used throughout this paper.  We adopt a $H_0 = 70$
km s$^{-1}$ Mpc$^{-1}$, $\Omega_M = 0.3$, $\Omega _{\Lambda} = 0.7$
cosmology.

\section {Observations}
\label{observations.sec}

\subsection {The GOODS Chandra Deep Field South}
\label{cdfs.sec}
Centered on $(\alpha,\delta)$ = (03:32:30, -27:48:30), the CDFS
(Giaconni et al. 2000) has been targeted by most of today's major
telescope facilities, both in imaging mode over the whole spectral
range and in spectroscopic mode.  Among the surveys carried out in
areas included in the CDFS are K20 (Cimatti et al. 2002), GMASS
(Cimatti et al. in preparation), VVDS (Le F\`{e}vre et al. 2004), and
the Hubble Ultra Deep Field (Beckwith et al. 2006).  In this section,
we describe the public GOODS-South dataset that we used to build a
$K_s$-band selected catalog containing homogeneous colors from the
optical to 24 $\mu$m.

\subsection {The WFI $U_{38}BVRI$ data}
\label{DPS.sec}
Ground-based optical imaging of the Extended CDFS with the Wide Field
Imager (WFI) on the ESO/MPG 2.2m telescope was obtained as part of the
COMBO-17 (Wolf et al. 2004) and ESO DPS (Arnouts et al. 2001) surveys.
We use the central part of the WFI $U_{38}$, $B$, $V$, $R$, and $I$
imaging, where it overlaps with the GOODS $K_s$-band imaging, that has
been re-reduced by the GaBoDS consortium (Erben et al. 2005;
Hildebrandt et al. 2006).  The images were released on a $0\farcs 238$
pixel$^{-1}$ scale and have exposure times of 49, 69, 105, 88, and 35
ks in $U_{38}$, $B$, $V$, $R$, and $I$ respectively.

\subsection {The ACS $B_{435}V_{606}i_{775}z_{850}$ data}
\label{ACS.sec}
During 5 epochs of observations, the ACS camera on HST acquired
imaging of the GOODS-South field in 4 filter bands: F435W, F606W,
F775W, and F850LP (hereafter referred to as
$B_{435}$,$V_{606}$,$i_{775}$, and $z_{850}$).  Exposure times
amounted to 2, 1.7, 1.7, and 3.3 hours respectively.  The mosaics
(version v1.0; Giavalisco et al. 2004), were drizzled onto a pixel
scale of $0\farcs 03$ pixel$^{-1}$.  From the 150 arcmin$^2$ area that
is well covered by the $K_s$-band detection image, 138 arcmin$^2$ is
well exposed with ACS.  We restrict our analysis of the total IR
properties of the distant galaxy population in
\S\ref{science.sec} to this overlap region.

\subsection{The ISAAC $JHK_s$ data}
\label{ISAAC.sec}
We use the ESO/GOODS data release v1.5 \\
(http://www.eso.org/science/goods/releases/20050930/) to complement
the optical observations with NIR imaging by the Very Large Telescope
(VLT).  For a full description of the dataset, we refer to Vandame et
al. (in preparation).  Briefly, the v1.5 data release consists of 24
fully reduced VLT/ISAAC fields in the $J$ and $K_s$ bands and 19 fields
in the $H$-band, each with a $2\farcm 5 \times 2\farcm 5$ FOV and $0\farcs
15$ pixel$^{-1}$ scale.  The ISAAC data were reduced using the ESO/MVM
image processing pipeline (v1.9, see Vandame 2002 for the description
of an earlier version).  Exposure times varied from field to field,
with typical exposures of 3.2, 4.2, and 5 hours in $J$, $H$, and
$K_s$-band respectively, and respective ranges between ISAAC fields of
2-5 hours, 2-6 hours, and 3.6-7.5 hours.  The variations in depth resulting
from the unequal exposure times are discussed in
\S\ref{bg_limdepth.sec}.  A total area of 113 arcmin$^2$ is well
exposed in all optical and NIR filter bands.  Without a restriction on
the $H$-band, the covered area increases to 138 arcmin$^2$.

\subsection{The IRAC 3.6-8.0 $\mu$m data}
\label{IRAC.sec}
As a Spitzer Space Telescope Legacy Program, superdeep images of the
GOODS-South field were taken with the Infrared Array Camera (IRAC,
Fazio et al. 2004) on-board Spitzer.  Over 2 epochs the whole field
was covered in the 3.6 $\mu$m, 4.5 $\mu$m, 5.8 $\mu$m, and 8.0 $\mu$m
bands.  For each epoch, exposure times per channel per sky pointing
amounted to 23 hours.  With the telescope orientation being rotated by
180 degrees between the two epochs, the second epoch IRAC channel 1
and 3 observations targeted the area covered by IRAC channel 2 and 4
during the first epoch, and vice versa.  An overlap region of roughly
40 arcmin$^2$, including the Hubble Ultra Deep Field (Beckwith et
al. 2003), got twice the exposure time.  We use the data releases DR2
and DR3 for the second and first epoch respectively.  Images were
released on a $0\farcs 60$ pixel$^{-1}$ scale.  A full description of
the observations and reduction will be presented by Dickinson et
al. (in preparation).

\subsection{The MIPS 24 $\mu$m data}
\label{MIPS.sec}
The GOODS-South field was observed at 24 $\mu$m with the Multiband
Imaging Photometry for Spitzer (MIPS, Rieke et al. 2004) on-board
Spitzer, closely overlapping the IRAC fields with a position angle
that is rotated with respect to the IRAC observations by approximately
3 degrees.  The MIPS campaign led to a nearly uniform exposure time of
10 hours.  We use the version v0.30 reduced images, released on a
$1\farcs 20$ pixel$^{-1}$ scale, based on the Spitzer Science Center
(SSC) Basic Calibrated Data (BCD) pipeline (version S11.0.2).

\section {Final images}
\label{finalim.sec}
In this section, we describe the image quality of the publicly
released data products, the subsequent steps undertaken to obtain the
final mosaics from which the photometric catalog is extracted, and the
limiting depths reached at all wavelengths.  The main characteristics
of the multi-wavelength data set are summarized in Table\ \ref{dataset.tab}.

\begin{deluxetable*}{llrrr}
\tablecolumns{5}
\tablewidth{0pc}
\tablecaption{Summary of the multi-wavelength data set for the GOODS-South field \label{dataset.tab}}
\tablehead{
\colhead{Camera} & \colhead{Filter} & \colhead{$\lambda _c$} & \colhead {FWHM\tablenotemark{a}} & \colhead {3$\sigma$ Total Limiting Magnitude} \\
 & & (\AA) & (arcsec) & (AB mag) \\
}
\startdata
\multicolumn{5}{c}{Ground-based Optical Bands} \\
\cline{1-5}
\multicolumn{5}{c}{} \\
WFI	& $U_{38}$	& 3656	& 0.93	& 26.06 \\
WFI	& $B$		& 4601	& 0.97	& 27.21 \\
WFI	& $V$		& 5379	& 0.90	& 26.88 \\
WFI	& $R$		& 6516	& 0.78	& 26.99 \\
WFI	& $I$		& 8658	& 0.93	& 25.00 \\
\cline{1-5}
\multicolumn{5}{c}{} \\
\multicolumn{5}{c}{Space-based Optical Bands} \\
\cline{1-5}
\multicolumn{5}{c}{} \\
ACS	& $B_{435}$	& 4327	& 0.22	& 27.68 \\
ACS	& $V_{606}$	& 5957	& 0.22	& 27.81 \\
ACS	& $i_{775}$	& 7705	& 0.21	& 27.26 \\
ACS	& $z_{850}$	& 9072	& 0.22	& 26.90 \\
\cline{1-5}
\multicolumn{5}{c}{} \\
\multicolumn{5}{c}{Near-infrared Bands} \\
\cline{1-5}
\multicolumn{5}{c}{} \\
\vspace{0.2in}
ISAAC	& $J$		& 12379	& 0.38 - 0.61	& 25.82 - 26.42 \\
ISAAC	& $H$		& 16517	& 0.35 - 0.65	& 25.39 - 25.95 \\
ISAAC	& $K_s$		& 21681	& 0.38 - 0.58	& 25.02 - 25.96 \\
\cline{1-5}
\multicolumn{5}{c}{} \\
\multicolumn{5}{c}{Mid-infrared Bands} \\
\cline{1-5}
\multicolumn{5}{c}{} \\
IRAC	& ch1		& 35634	& 1.6	& 26.15 \\
IRAC	& ch2		& 45110	& 1.7	& 25.66 \\
IRAC	& ch3		& 57593	& 1.9	& 23.79 \\
IRAC	& ch4		& 79595	& 2.0	& 23.70 \\
MIPS	& 24 $\mu$m	& 237000 & 6.0	& 21.30
\enddata
\tablenotetext{a}{The intrinsic ACS PSF has a FWHM of $0\farcs 12$.  The values quoted in this table were measured on the $5 \times 5$ block averaged mosaics.  The ACS photometry was performed on the images convolved to the broadest NIR PSF.}
\end{deluxetable*}

\subsection {Pixel scales and large scale backgrounds}
\label{pixel.sec}

First, we converted the ACS images to the $0\farcs 15$ pixel$^{-1}$
scale of the $K_s$-band detection image by $5 \times 5$ pixel block
averaging with flux conservation.  All space-based optical and NIR
photometry was performed on this pixel scale using the SExtractor
software version 2.2.2 (Bertin \& Arnouts 1996) (see
\S\ref{det_phot.sec}).

A source fitting algorithm developed by Labb\'{e} et al. (in
preparation), especially suited for heavily confused images for which
a higher resolution prior (in this case the $K_s$-band image) is
available, was used to extract photometry from the WFI, IRAC and MIPS
images (see \S\ref{MIR_phot.sec}).  The WFI images were registered to
the $K_s$-band mosaic with a $0\farcs 15$ pixel$^{-1}$ scale using
position measurements of stars and compact objects and the IRAF geomap
and geotran tasks.  For the MIR photometry, the algorithm requires a
higher resolution image than provided by the IRAC and MIPS images.
However, the native $K_s$-band pixel scale makes it more
computationally expensive without benefit in accuracy.  A $2 \times 2$
block averaged version of the $K_s$-band mosaic was therefore produced
with a $0\farcs 3$ pixel$^{-1}$ scale.  We registered the publicly
released IRAC and MIPS images onto this $K_s$-band image using the WCS
information in the image header in combination with a minor additional
shift based on position measurements of stars and compact sources,
again forcing flux conservation.  We note that the source fitting
algorithm takes care of residual shifts.  Since the programme does not
take into account large scale background variations, these were
removed a priori from the IRAC and MIPS images by subtracting
SExtractor background images produced with large background mesh
settings (BACK\_SIZE=64, BACK\_FILTERSIZE=3 on a $0\farcs 3$
pixel$^{-1}$ scale).  The subtracted background varied over the image
by -2 to 0 times the rms noise.

\subsection {Image quality and PSF matching}
\label{PSFmatch.sec}

\subsubsection {Space-based optical and NIR wavelengths}
\label{opt_NIR_PSF.sec}
In order to obtain consistent color measurements, we match all
space-based optical and NIR images to a common resolution, namely that
of the field with the broadest point spread function (PSF).  Since the
ground-based optical, IRAC, and MIPS images have a significantly
broader PSF, they are not included in the simple PSF matching, but are
treated separately by a source-fitting algorithm (see
\S\ref{MIR_phot.sec}).  In this section, we describe the selection of
stars used to build the PSFs, the construction of the PSFs for the ACS
$B_{435}$-, $V_{606}$-, $i_{775}$-, and $z_{\rm
850}$-mosaics and for each of the ISAAC fields in the $J$-, $H$-, and
$K_s$-band, the construction of the convolving kernels, and the
quality of the PSF matching.

First, we compiled a list of bright, isolated, unsaturated stars.
Initially, well-exposed objects with $(J-K_s)_{\rm AB} < 0.04$ and
$K_{s,{\rm AB}}^{\rm tot}<22.86$ mag were selected from a preliminary
catalog of the CDFS.  For ISAAC fields where the number of $J-K_s$
selected stars was low, we complemented the sample with stars from the
EIS stellar catalog (Groenewegen et al. 2002).  During a first
iteration, the list was cleaned from galaxy-like objects, stars with
neighbors within 3'' radius, stars too close to the edge of an image,
objects that were not identified in the ACS r1.1z catalog (Giavalisco
et al. 2004) or resolved objects as defined by having a FWHM in the
$z_{850}$-band larger than $0\farcs 13$.  Measurements of the FWHM
were performed by fitting Moffat profiles to the stars using the
imexam task in IRAF.  Finally, we inspect by eye the radial profiles
and curves of growth, produced with the IRAF tasks radprof and phot
respectively.  The final list contained 3 to 5 stars per ISAAC field,
with the exception of field f30, for which only 1 good star was
available.  The numbers of stars used to build the ACS PSFs were 31,
45, 49 and 53 for the $B_{435}$, $V_{606}$, $i_{775}$, and
$z_{850}$ mosaics respectively.

It is sufficient to construct one PSF image per band for the whole ACS
mosaic since for space-based imaging the PSF is uniform across all
pointings.  The ground-based data of each ISAAC field were taken under
different average seeing conditions, hence the need for the
construction of an individual PSF image for each field and each NIR
band.  We computed the $J$-, $H$-, and $K_s$-band PSF images per ISAAC
field by averaging the registered and flux-normalized images of the
selected stars.  The flux was normalized within $1\farcs 5$ diameter
apertures rather than the total aperture to optimize the
signal-to-noise and avoid contributions from residual neighboring
sources.  Any neighbors in the ISAAC images of good stars,
sufficiently far not to bias the FWHMs and PSFs, were masked while
averaging.  This method was preferred over taking the median, since
only a handful of good stars per ISAAC field were available to build a
PSF.  PSFs for the ACS mosaics were constructed from a large enough
number of stars to average out any influence of faint neighboring
sources without masking.

\begin {figure} [t]
\centering
\plotone{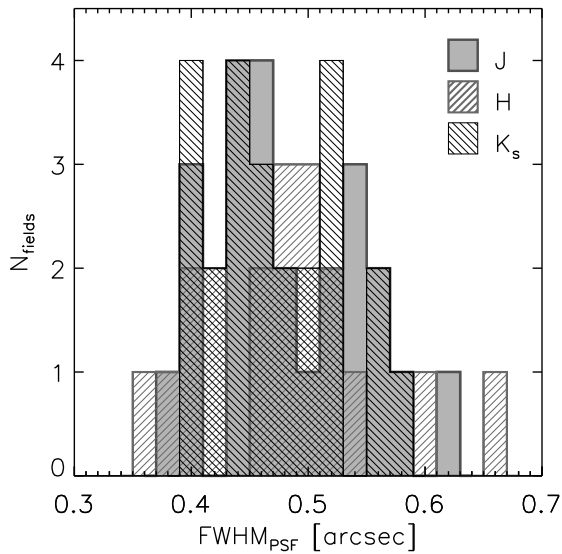}
\caption{\small Distributions of seeing FWHM for the ISAAC $J$, $H$, and
$K_s$ observations.  Moffat profiles were fitted to the PSFs that were
built from bright, isolated, unsaturated stars for each field and band
separately.
\label {FWHMhist.fig}
}
\end {figure}

The ACS PSFs in the $B_{435}$-, $V_{606}$-, $i_{775}$-, and
$z_{850}$-band on the $5 \times 5$ block-averaged ACS mosaics are well
fit by Moffat profiles with $\beta=3.8$, 4.0, 3.7, and 3.6, and a FWHM
of $0\farcs 22$, $0\farcs 22$, $0\farcs 21$, and $0\farcs 22$
respectively.  The FWHM of the NIR ISAAC observations varied from
$0\farcs 35$ to $0\farcs 65$, with median values of $0\farcs 47$,
$0\farcs 48$, and $0\farcs 47$ in $J$, $H$, and $K_s$ respectively.
The values of the Moffat $\beta$ parameter were 3.1, 2.9, and 3.0 in
$J$, $H$, and $K_s$.  Fig.\
\ref{FWHMhist.fig} illustrates the distribution of FWHMs of the
natural PSFs for the individual ISAAC fields.  In all of the
considered bands and fields, the FWHM of the individual stars were
within $\approx$10\% of that measured on the final average PSF of the
image.  We adopted the $0\farcs 65$ $H$-band PSF of ISAAC field f15 as
target to which all higher resolution images were matched.

We computed the kernel for convolution for each ISAAC field and band
separately, using the Lucy-Richardson deconvolution algorithm.  The
kernels are well represented by a Moffat profile with $\beta$
parameters varying from field to field and band to band.  The ratio of
the growth curve of the convolved PSF over that of the target PSF is a
good measure for the PSF matching accuracy.  Overall, the ratio of
growth curves deviated by at most 5.1\% from unity for apertures
between 1'' ($\approx$1.5FWHM of the PSF of the smoothed field maps)
and 6'' (the reference aperture for photometric calibration), with an
average of $0.54\% \pm 0.90\%$.  Flux within 3'' radius is well
conserved during the convolution process, with a median deviation of
0.06\% and a median absolute deviation of 0.6\%.

The construction of convolving kernels for the ACS mosaics required an
extra step.  The kernels obtained from deconvolution with the IRAF
Lucy task had significant noise in the outer parts, leading to noise
spikes around bright point-like sources in the convolved ACS mosaics.
To remove these artifacts, we modeled the ACS-to-ISAAC kernels by
fitting isophotes using the IRAF tasks ellipse and bmodel, and used
the modeled kernels for the convolution.  This is possible because the
kernels are otherwise well behaved and very azimutally symmetric.  The
ACS-to-ISAAC kernels have a nearly Moffat profile with a $\beta$
parameter of 3.

Because of the different basic shapes of the ACS and ISAAC PSFs, an
excellent matching over the relevant radii is more difficult than in
general among ISAAC fields.  Nevertheless, the offsets of the growth
curve ratios between 1'' and 6'' are limited to below 4.7\%, with an
average of $1.58\% \pm 1.32\%$.  The average over all stars of the
ratio of the flux measured within 3'' radius in the convolved and that
measured in the natural image showed a flux conserving accuracy of
0.7\% or better for all ACS bands.

\subsubsection {Ground-based optical and MIR wavelengths}
\label{MIR_PSF.sec}
The seeing PSF of the WFI observations is significantly broader than
that of our $K_s$-band detection image ($\sim 0.9$'' FWHM).  The FWHM
measured on the average image of bright, isolated stars in the IRAC
images even amounts to $1\farcs 6$, $1\farcs 7$, $1\farcs 9$, and
$2\farcs 0$ for the 3.6 $\mu$m, 4.5 $\mu$m, 5.8 $\mu$m, and 8.0 $\mu$m
bands respectively.  The MIPS 24 $\mu$m beam has a FWHM as large as
6''.  Since confusion and blending effects are unavoidable in deep
observations at this resolution, we decide not to degrade the
space-based optical and NIR images to the MIR resolution.  Instead, we
construct PSFs and convolving kernels similarly as described in
\S\ref{opt_NIR_PSF.sec}, but apply them using a source fitting
algorithm that makes fully use of the higher resolution information in
the $K_s$-band detection image (see
\S\ref{MIR_phot.sec}, Labb\'{e} et al. in preparation).

\subsection {Zero points}
\label{zeropoints.sec}
The zero-point calibrations for all bands but the $H$-band were taken
from the respective GOODS data release.  In the case of the NIR ISAAC
observations, the publicly released zero points were based on SOFI
images of the EIS-DEEP and DPS infrared surveys conducted over the
same region (Vandame et al. 2001), which themselves were
photometrically calibrated using standard stars from Persson et
al. (1998).  That procedure yielded zero points with rms scatters
ranging between 0.01 and 0.06 mag in the $J$-band, 0.01 and 0.08 mag in
the $K_s$-band and up to 0.17 mag in the $H$-band.

\begin{deluxetable}{lc}
\tablecolumns{2}
\tablewidth{0pc}
\tablecaption{H-band zero points in the AB system derived from the NIR stellar locus \label{zp.tab}
}
\tablehead{
\colhead{Field} & \colhead{$H$-band zero point}
}
\startdata
03 & 25.99 \\
04 & 26.02 \\
05 & 26.07 \\
08 & 25.89 \\
09 & 25.92 \\
10 & 25.94 \\
11 & 25.93 \\
13 & 26.02 \\
14 & 25.82 \\
15 & 26.07 \\
16 & 25.97 \\
19 & 25.86 \\
20 & 25.89 \\
21 & 25.97 \\
22 & 26.03 \\
23 & 26.07 \\
24 & 25.95 \\
25n & 25.94 \\
26n & 26.09
\enddata
\end{deluxetable}

To improve on the $H$-band calibration, we make use of stellar
photometry in the FIRES HDFS (L03) and \1054 (FS06) fields, for which
$H$-band zero points were determined to a $\sim 0.03$ mag accuracy.  For
each of the 19 ISAAC fields with $H$-band coverage, we measured the mean
offset of the stars used for PSF matching along the $J-H$ axis of a
$J-K_s$ versus $J-H$ color-color diagram with respect to the stellar
locus in the FIRES fields.  Assuming the $J$- and $K_s$-bands are well
calibrated, this immediately provides us with the $H$-band zero point
corrections to be applied.  We list the derived $H$-band zero points in
Table\ \ref{zp.tab}.  Zero-point corrections ranged from -0.18 mag to
0.09 mag, with a median correction over all fields of -0.03 mag.
After applying the zero-point correction, the median absolute deviation
in $J-H$ color of individual stars around the stellar locus is 0.03 mag,
similar as measured for the FIRES HDFS and \1054 fields.

\subsection {Mosaicing and astrometry}
\label{astrometry.sec}
Here, we describe the combination of the smoothed ISAAC NIR fields and
the astrometric precision of the final mosaics.  The $5 \times 5$
blocked and smoothed ACS $i_{775}$-band mosaic was adopted as astrometric
reference image.  The astrometric solution for the ACS data itself was
based on a cross-identification of sources with deep ground-based WFI
data that on its turn was astrometrically matched to stellar positions
in the Guide Star Catalog 2 (GSC2, STScI 2001).  The final solution
had a $2\sigma$-clipped rms deviation of $\lesssim 0\farcs 01$ in
ACS-to-ACS and $0\farcs 12$ in ACS-to-ground difference.

\begin {figure} [t]
\centering
\plotone{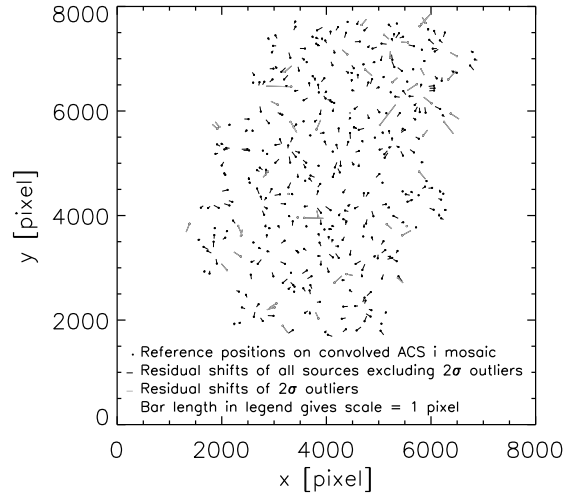}
\caption{\small Map of residual shifts of compact sources in the
$K_s$-band mosaic with respect to the reference ACS $i_{775}$-band mosaic.
$2\sigma$-clipped reference sources used for the alignment are
indicated in black.  Grey vectors represent the residual shift of the
$2\sigma$ outliers.
\label {astrometry.fig}
}
\end {figure}

The smoothed ISAAC fields were registered onto the smoothed ACS $i_{775}$-band
mosaic by applying simple x- and y- shifts without further distortion
corrections.  In each ISAAC field, we measured the shift with respect
to the ACS $i_{775}$-band mosaic for stars and compact sources using the
imexam task in IRAF.  A $2\sigma$-clipped sample of reference sources
typically consisted of 15-20 objects per ISAAC field.  The
$2\sigma$-clipping was applied to exclude sources with strong
morphological k-corrections.  The difference between the shifts
implied by individual reference sources and the final astrometric
solution had a standard deviation of less than $0\farcs 09$ in all NIR
bands.  A map of residual shifts for the $K_s$-band mosaic with
respect to the convolved ACS $i_{775}$-band mosaic is presented in Fig.\
\ref{astrometry.fig}.

First we applied the fractional pixel shift for each ISAAC field in
each band using the IRAF imshift task with a cubic spline
interpolation.  Next, we summed the integer pixel shifted fields
applying an identical weighting scheme as described by FS06 to optimize
the S/N for point sources, namely:
\begin {equation}
w_{\rm pix} = \frac{w_{\rm norm}}{(rms_{\rm 1.5FWHM})^2}
\label{weight.eq}
\end {equation}
where the weight factor for a given pixel $w_{\rm pix}$ equals its
value in the normalized weight map $w_{\rm norm}$, scaled with the
square of the rms noise measured within apertures of 1.5FWHM diameter
that are placed randomly in empty regions of the ISAAC field (see
\S\ref{bg_limdepth.sec}).  Here, $w_{\rm norm}$ is simply the weight
map as released by ESO/GOODS normalized to a maximum of unity.  It
thus contains information on the relative integration time per pixel
and accounts for bad/hot pixels and pixels with other artifacts that
were excluded.  The weight maps for each field given by Eq.\
\ref{weight.eq} were combined into a mosaic that accounts for the
effective background noise over the entire area.  In the catalog, we
give the weight value for each object measured on this effective
weight map, normalized to the median weight value in the considered
band.

The standard deviation of positional offsets between reference sources
on the registered WFI images and the $K_s$-band mosaic were less than
$0\farcs 1$.

We chose not to combine the 2 epochs of IRAC observations into one
mosaic because the 180 degrees difference in position angle would lead
to a different PSF shape in the overlap region than in either of both
single epoch areas, demanding the use of a different convolving kernel
over different parts of the field.  Instead, we treat each of the IRAC
epochs independently, providing an empirical quality check of the
photometry in the overlap region.  The registration of each of the
IRAC images (epoch 1 and 2) onto the 2x2 blocked $K_s$-band image has
a positional accuracy of better than $0 \farcs 4$, as measured from
offsets between bright star positions on the IRAC and $K_s$-band images.
The positional accuracy for the MIPS images is of the order of $0
\farcs 3$ rms.  We note that minor positional offsets between the
$K_s$ and IRAC/MIPS image are solved for by the source fitting
algorithm applied to IRAC and MIPS photometry (see
\S\ref{MIR_phot.sec}).

\subsection {Signal to noise and limiting depths}
\label{bg_limdepth.sec}

\begin {figure} [htbp]
\centering
\plotone{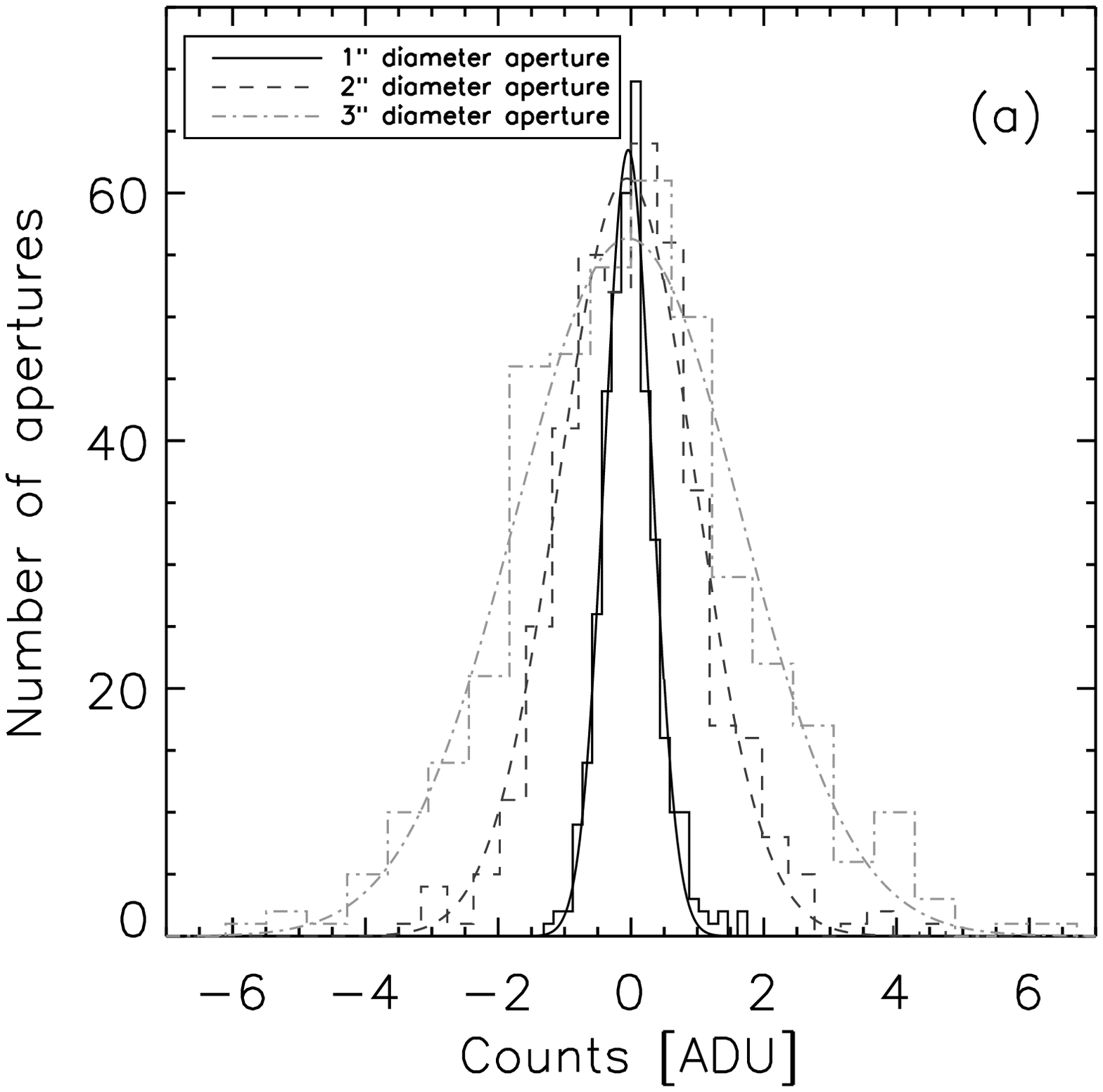}
\vspace{0.1in}
\plotone{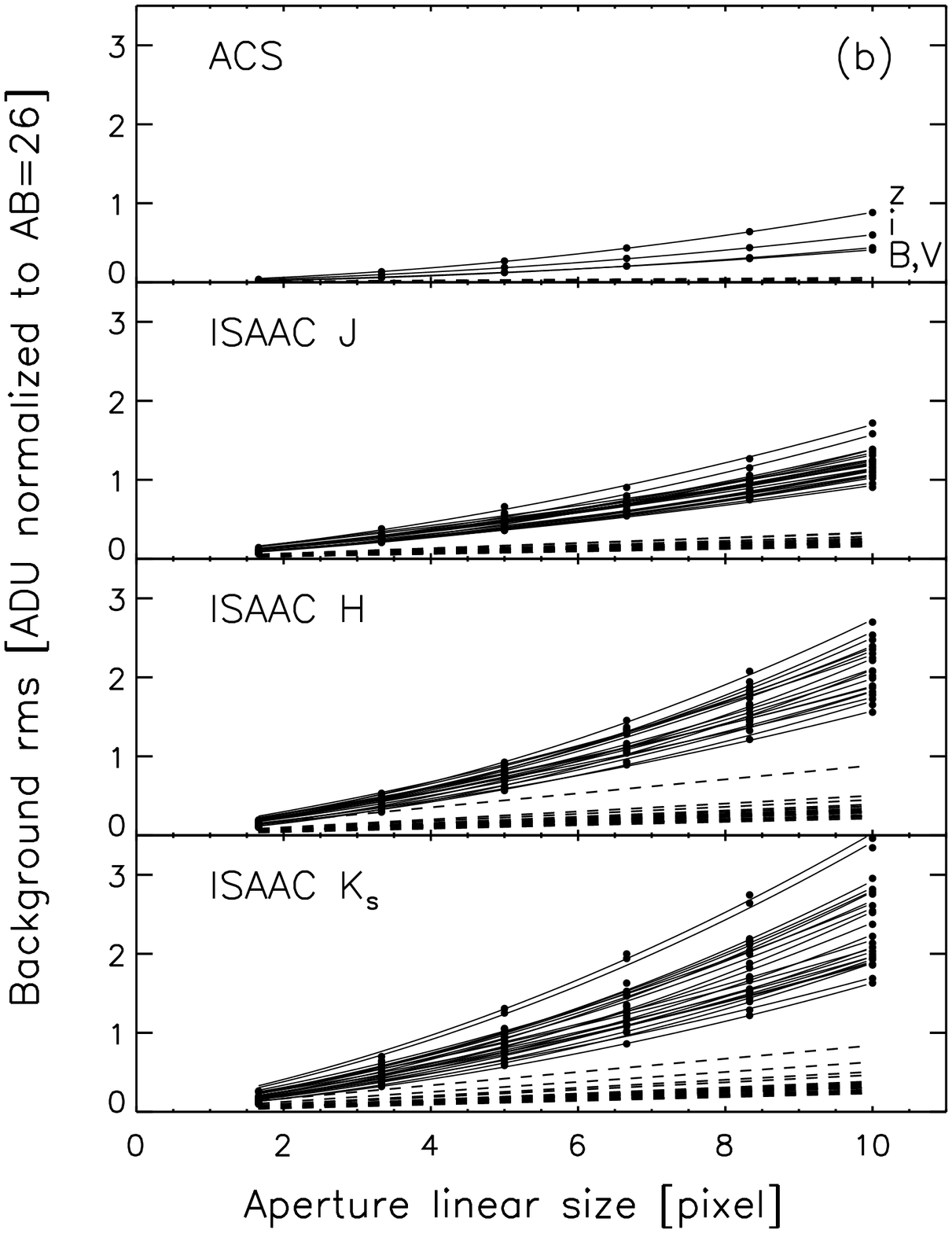}
\caption{\small The background rms derived from the distribution of
fluxes within empty apertures.  (a) Distribution of empty aperture
fluxes within a 1", 2", and 3" aperture diameter on the $K_s$-band
image of ISAAC field f15.  The distribution is well described by a
Gaussian with an increasing width for increasing aperture size.  (b)
Background rms as derived from flux measurements within empty
apertures versus aperture size for the ACS bands and the $J$, $H$, and
$K_s$ ISAAC fields.  Solid lines represent the functional form from
Eq.\ \ref{error.eq} fit to the observed rms noise values.  Dashed
lines indicate a linear extrapolation of the pixel-to-pixel rms.
Correlations between pixels introduce a stronger than linear scaling
with aperture size.
\label {limdepth.fig}
}
\end {figure}

We analyzed the noise properties of the optical-to-24 $\mu$m
imaging following the same approach as for the FIRES HDFS (L03) and
\1054 (FS06) data.  Briefly, the technique uses aperture photometry on
empty parts of the image to quantify the rms of background pixels
within the considered aperture size.  For each convolved ISAAC field
in each band, between 200 and 400 non-overlapping apertures were
randomly placed at a minimum distance of 5'' from the nearest
segmentation pixels in a SExtractor segmentation map.  For a given
aperture size, the distribution of empty aperture fluxes is
well-fitted by a Gaussian, as illustrated in Fig.\
\ref{limdepth.fig}(a).  We applied a $5\sigma$ clipping in determining
the background rms.  Panel (b) of Fig.\ \ref{limdepth.fig} shows that
a simple linear scaling of the measured background rms
$\sigma(N)=N\bar{\sigma}$ , where $N=\sqrt{A}$ is the linear size of
the aperture with area A and $\bar{\sigma}$ is the pixel-to-pixel rms,
would lead to underestimated flux uncertainties.  The reason is that
correlations between neighboring pixels were introduced during the
reduction and PSF matching.  We model the background rms as a function
of aperture size with a polynomial of the form
\begin {equation}
\sigma_i(N) = \frac{N\bar{\sigma}(a_i + b_i N)}{\sqrt{w_i}}
\label{error.eq}
\end {equation}
where $i$ refers to the considered band and field, and the weight term
$w_i$ is derived from the weight map of the respective field.  Fig.\
\ref{limdepth.fig}(b) illustrates the variations in depth for the
different ISAAC fields, originating from variable integration times
and observing conditions, and reflected in the range of flux
uncertainties for objects with similar color aperture in the final
catalog.  For example, the upper two curves in the ISAAC $K_s$ panel
correspond to fields f03 and f04 that had the lowest integration time.

For the ACS mosaics, we used the same empty apertures as for the NIR,
provided they were within the ACS FOV.  Every object below the
$K_s$-band detection threshold, even though detectable in the ACS
imaging, contributes to the background noise and photometric
uncertainties of $K_s$-band detected sources.  If we were to restrict
our empty aperture analysis to apertures that contain neither
$K_s$-band nor ACS segmentation pixels, the background rms estimates
for the ACS mosaics would decrease by 3 to 9\%.  In Fig.\
\ref{limdepth.fig}(b), we scaled the background rms measured on the
ACS and ISAAC images to the flux corresponding to AB = 26.

Uncertainties on the fluxes were assigned based on these empirical
noise measurements, for an aperture of the same area as for the
photometric aperture of each object.

Limiting depths of the WFI, IRAC, and MIPS images were measured with
the same empty aperture method.  The total limiting AB magnitudes
(3$\sigma$ for point sources, i.e., corrected for the flux missed
outside the aperture used for photometry) are listed in Table\
\ref{dataset.tab}.

\section {Source detection and photometry}
\label{det_phot.sec}

\subsection {$K_s$-band Detection}
\label{detection.sec}
We aimed to construct a catalog that is especially suited to extract
stellar mass-limited samples from (e.g., van Dokkum et al. 2006).
Although the rest-frame NIR, probed by IRAC, is a better tracer for
stellar mass than the rest-frame optical, the downside is its coarser
resolution, leading to severe confusion.  Therefore, we decided to
detect sources in the observed $K_s$-band.

We used the SExtractor v2.2.2 source extraction software by Bertin \&
Arnouts (1996) to detect sources with at least 1 pixel above a surface
brightness threshold of $\mu (K_{s,{\rm AB}}) = 24.6$ mag arcsec$^{-2}$,
corresponding to $\approx 5\sigma$ of the rms background for a typical
$K_s$-band field.  Setting the threshold to the same number of ADUs
across the image instead of adopting a $S/N$ criterion was favored,
since in the latter case the varying noise properties in the
$K_s$-band mosaic would lead to different limiting magnitudes and
limiting surface brightnesses from one field to the other.  We
smoothed the detection map with a gaussian filter of $FWHM = 0\farcs
65$, the size of the PSF in the detection image.  This procedure
optimizes the detection of point sources.

The resulting catalog contains 6308 sources, 5687 of which have a
position on the $K_s$-band mosaic where the pixel weight is larger
than 30\% of the median weight.  We recommend applying such a weight
criterion in order to construct samples with robust photometry.
Running SExtractor with identical parameters on the detection map
multiplied by -1, we obtain a total of 43 spurious sources in the area
with more than 30\% of the median weight.  Only one of these has
$S/N_{K_s} > 5$.

SExtractor flagged 12\% of the detected sources as blended and/or
biased.  These sources were treated separately in the photometry.

\subsection {Photometry}
\label{photometry.sec}

\subsubsection {Space-based optical and NIR photometry}
\label{opt_NIR_phot.sec}
We performed the photometry on the convolved ACS $B_{435}V_{\rm
606}i_{775}z_{850}$ and ISAAC $JHK_s$ mosaics using SExtractor
in dual image mode, with the $K_s$-band mosaic as detection map.  We
derive the color and total aperture from the detection image.  The
same apertures were used in each band.  We follow L03 and FS06 in
defining the color aperture based on the $K_s$-band isophotal
aperture, more precisely on the equivalent circularized isophotal
diameter $d_{\rm iso} = 2(A_{\rm iso}/\pi)^{1/2}$, where $A_{\rm iso}$ is the area
of the isophotal aperture.  For isolated sources, we apply
\begin {equation}
{\small
APER(COLOR)=\left\{\begin{array}{cl}
	       \vspace{0.2cm}
               APER(ISO),       & 1\farcs 0 < d_{\rm iso} < 2\farcs 0 \\
	       \vspace{0.2cm}
	       APER(1\farcs 0), & d_{\rm iso} \leq 1\farcs 0 \\
	       APER(2\farcs 0), & d_{\rm iso} \geq 2\farcs 0 
            \end{array}\right.
}
\label{col_ap_isolated.eq}
\end {equation}
where APER(ISO) refers to the isophotal aperture defined by the
surface brightness detection threshold.  Blended sources (indicated
with SExtractor flag ``blended'' or ``biased'') were treated
separately,
\begin {equation}
{\small
APER(COLOR)=\left\{\begin{array}{cl}
	       \vspace{0.2cm}
               APER(d_{\rm iso}/s),       & 1\farcs 0 < d_{\rm iso}/s < 2\farcs 0 \\
	       \vspace{0.2cm}
	       APER(1\farcs 0), & d_{\rm iso}/s \leq 1\farcs 0 \\
	       APER(2\farcs 0), & d_{\rm iso}/s \geq 2\farcs 0 
            \end{array}\right.
}
\label{col_ap_blended.eq}
\end {equation}
where the reduction factor s for the aperture sizes is introduced to
minimize contamination by close neighbors.  We adopt the optimal value
of $s=1.4$ that was determined from experimentation by L03 and FS06.

The motivation for the tailored isophotal apertures defined in Eq.\
\ref{col_ap_isolated.eq} and Eq.\ \ref{col_ap_blended.eq} is that it
maximizes the $S/N$ of the flux measurement.  The minimum diameter of
$1\farcs 0$ corresponds to 1.5 $\times$ FWHM of the PSF-matched
mosaics.  The maximum diameter of $2\farcs 0$ was adopted to avoid
flux from neighboring sources and avoid the large uncertainties
corresponding to large isophotal apertures.

SExtractor's ``MAG\_AUTO'' was used to derive the total flux of the
$K_s$-band detected objects, unless the source was blended, in which
case the total aperture was set to the color aperture:
\begin {equation}
{\small
APER(TOTAL)=\left\{\begin{array}{cl}
	       \vspace{0.2cm}
               APER(AUTO),       & isolated\ sources \\
	       APER(COLOR),      & blended\ sources \\
            \end{array}\right.
}
\label{tot_ap.eq}
\end {equation}
Finally, an aperture correction was applied to compute the total
integrated flux.  The correction factor equaled the ratio of the flux
within an aperture of radius 3'' and radius $r_{\rm tot}=(A_{\rm
tot}/\pi)^{1/2}$, based on the curve of growth of the PSF constructed
from stellar profiles.  Here, $A_{\rm tot}$ represents the area of the
total aperture.  The reference aperture of 3'' radius is sufficiently
large since less than 0.1\% of the flux for point sources falls
outside this radius.

Flux uncertainties in both color and total aperture were derived from
Eq.\ \ref{error.eq}.  The quoted uncertainties thus take into account
both the aperture size used for the flux measurement and the limiting
depth in the respective region of the mosaic.

We performed a large number of simulations to determine the accuracy
of our measurements of the total $K_s$-band magnitude.  In each
simulation, 200 artificial sources were added to the $K_s$-band
mosaic.  They had either exponential (Sersic n=1) or de Vaucouleurs
(Sersic n=4) light profiles that were truncated at 5 effective radii,
and random position angles and ellipticities.  Source magnitudes and
sizes were drawn from the true distribution in our catalog in order to
span the entire range of real galaxy properties.  Subsequently, we ran
SExtractor with an identical configuration as for the real catalog,
followed by the procedure to compute $K^{\rm tot}_{s,{\rm AB}}$ as
described above.  The recovered total magnitudes were within 0.1 (0.2)
mag from the input value for 89 (96)\% of the sources with exponential
light profiles.  Since our aperture corrections are based on the
stellar growth curve, our procedure could not account for all flux in
the extended wings of low surface brightness sources with a high
concentration (Sersic n=4).  The recovered total magnitudes were
within 0.1 (0.2) mag for 82 (91)\% of the sources with de Vaucouleurs
profiles.  The sources where the $K^{\rm tot}_{s,{\rm AB}}$
underestimate exceeded 0.2 mag generally had $\mu_e > 22.5$ mag
arcsec$^{-2}$.  Taking into account the spread in Sersic indices for
real sources in our catalog, we caution that underestimates of the
total $K_s$-band magnitude by over 0.1, 0.2, and 0.3 mag could occur
for roughly 13, 6, and 4\% of the galaxies in our catalog,
predominantly those with large size and low surface brightness.

\subsubsection {WFI, IRAC and MIPS 24 $\mu$m photometry}
\label{MIR_phot.sec}
In this section, we describe the photometry of $K_s$-band detected
objects in the ground-based optical (WFI) and Spitzer IRAC and MIPS 24
$\mu$m imaging of the CDFS.  For an in depth discussion of the source
fitting algorithm used, and simulations of its performance, we refer
the reader to Labb\'{e} et al. (in preparation).  A short description
with illustration was also presented by Wuyts et al. (2007).

Briefly, the information on position and extent of sources based on
the higher resolution $K_s$-band segmentation map was used to model
the lower resolution ground-based optical and 3.6 $\mu$m to 24 $\mu$m
images.  Each source was extracted separately from the $K_s$-band
image and, under the assumption of negligible morphological
k-corrections, convolved to the WFI, IRAC, respectively MIPS
resolution.  A fit to the WFI/IRAC/MIPS image was then made for all
sources simultaneously, where the fluxes of the objects were left as
free parameters.  Next, we subtracted the modeled light of neighboring
objects and measured the flux on the cleaned WFI/IRAC/MIPS maps within
a fixed aperture: 2'' for the WFI bands, 3'' for the IRAC bands and
6'' for the MIPS 24 $\mu$m band.  In order to compute a consistent
$K_s - $WFI/IRAC/MIPS color, we measured the source's flux
$f_{{\rm conv},K_s}$ on a cleaned $K_s$-band image convolved to the
WFI/IRAC/MIPS resolution within the same aperture.  We then scaled the
photometry to the same color apertures that were used for the
space-based optical and NIR photometry, allowing a straightforward
computation of colors over a $U_{38}$-to-8 $\mu$m wavelength baseline.
For the IRAC photometry, this means the catalog flux was computed as
follows:
\begin {equation}
f_{\rm IRAC,col} = f_{\rm IRAC,3''} * \frac{f_{K_s,{\rm col}}}{f_{{\rm conv},K_s,3''}}.
\end {equation}
For the WFI photometry, an identical procedure was followed.  An
aperture correction based on the growth curve of the 24 $\mu$m PSF was
applied to scale the 24 $\mu$m flux measurements to the integrated 24
$\mu$m flux under the assumption that the sources were point-like.

We note that several other authors developed similar algorithms to
reduce the effects of confusion (P\'{e}rez-Gonz\'{a}lez et al. 2005,
2008; G06a).  P\'{e}rez-Gonz\'{a}lez et al. (2005, 2008) compute
SExtractor photometry for an IRAC-selected catalog that is later
merged with shorter wavelength photometric catalogs.  For the IRAC
sources with multiple UV/optical/NIR counterparts within $2\farcs 5$,
the photometry was recomputed using a deconvolution method that is
similar to that by Labb\'{e} et al. (in preparation), except for two
aspects.  The algorithm by Labb\'{e} et al. (in preparation) takes
into account the spatial extent of the sources on the reference
($K_s$-band) image.  Moreover, Labb\'{e} et al. (in preparation) does
not adopt the best-fit flux from the source fitting as final
photometry, but rather measures the flux within an aperture on the
cleaned image (followed by an aperture correction).  This allows more
robust photometry in cases where the object profile varies from the
reference to the low resolution image.  A comparison with the
multi-wavelength photometry by G06a is presented in
\S\ref{MUSICphot.sec}.

\begin {figure*} [htbp]
\centering
\epsscale{1}
\plottwo{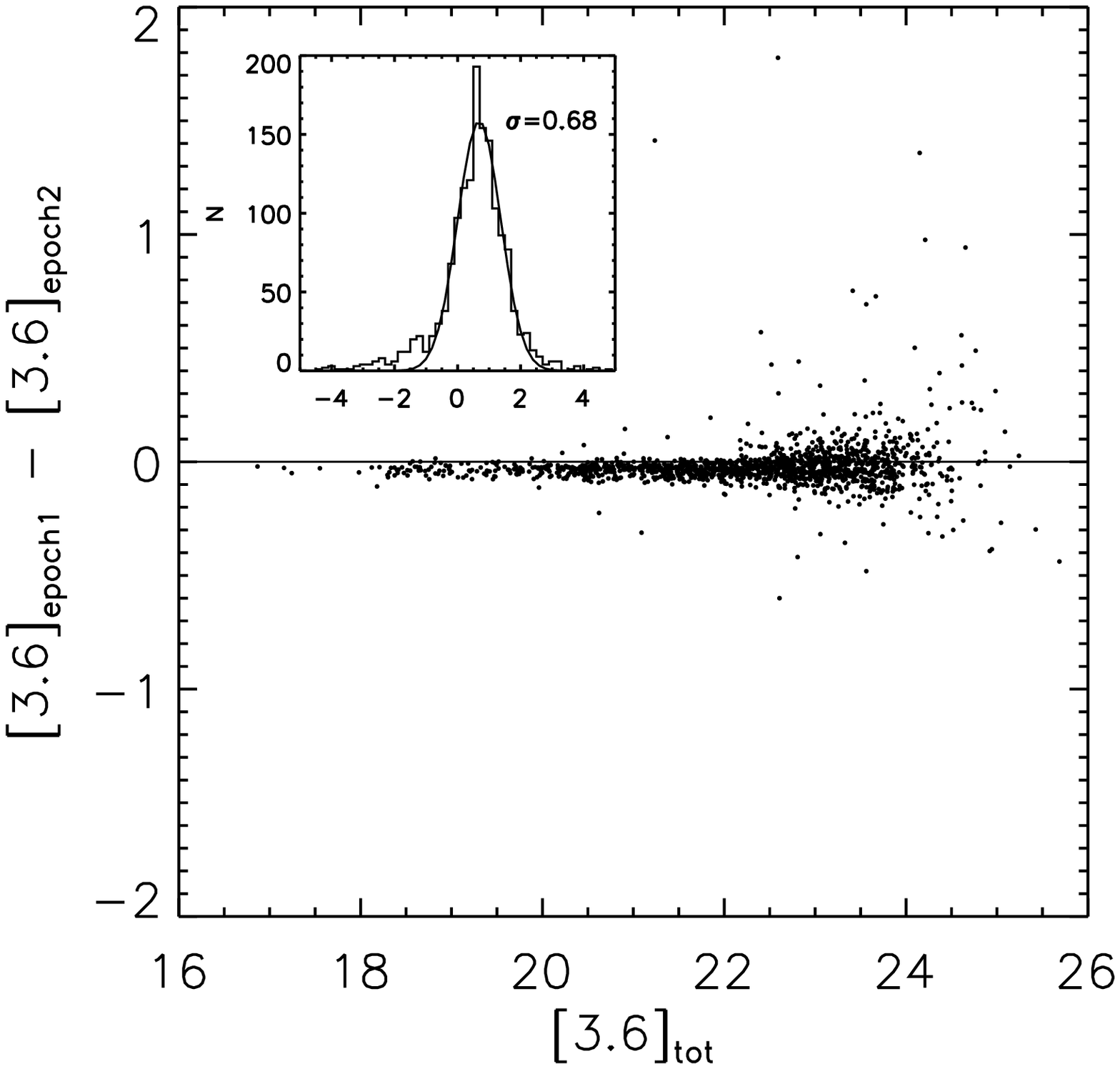}{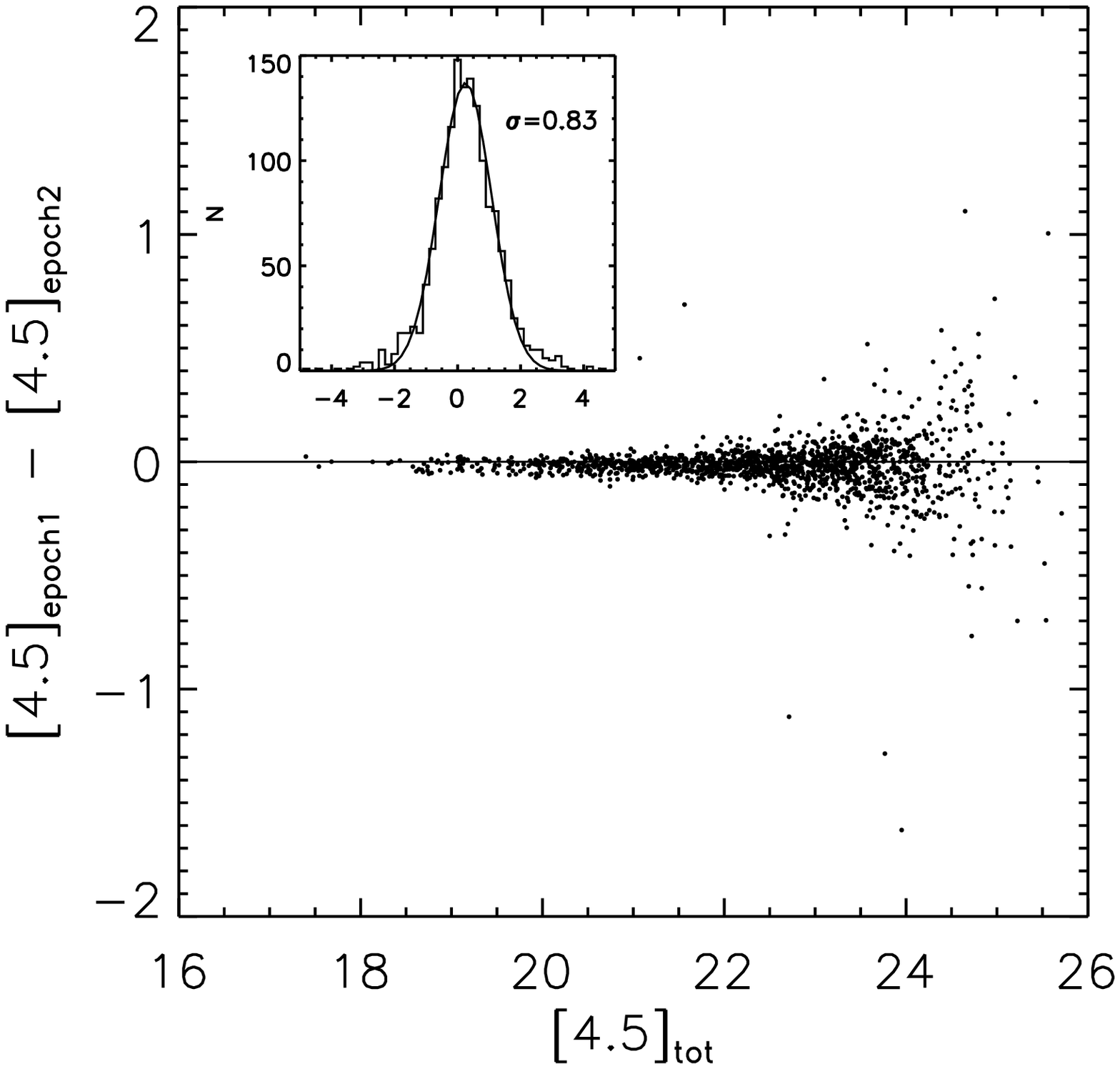}
\vspace{0.1in}
\plottwo{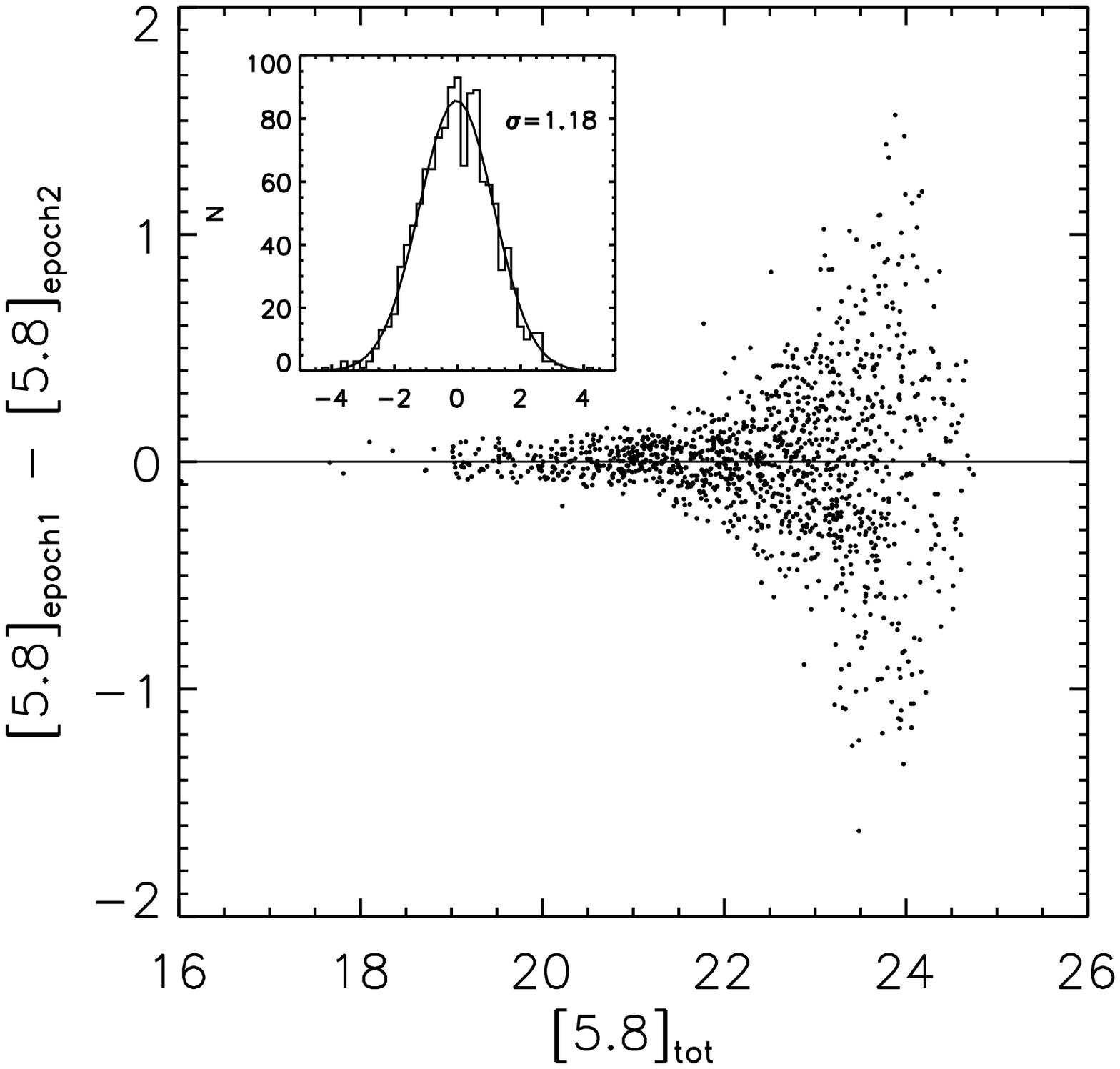}{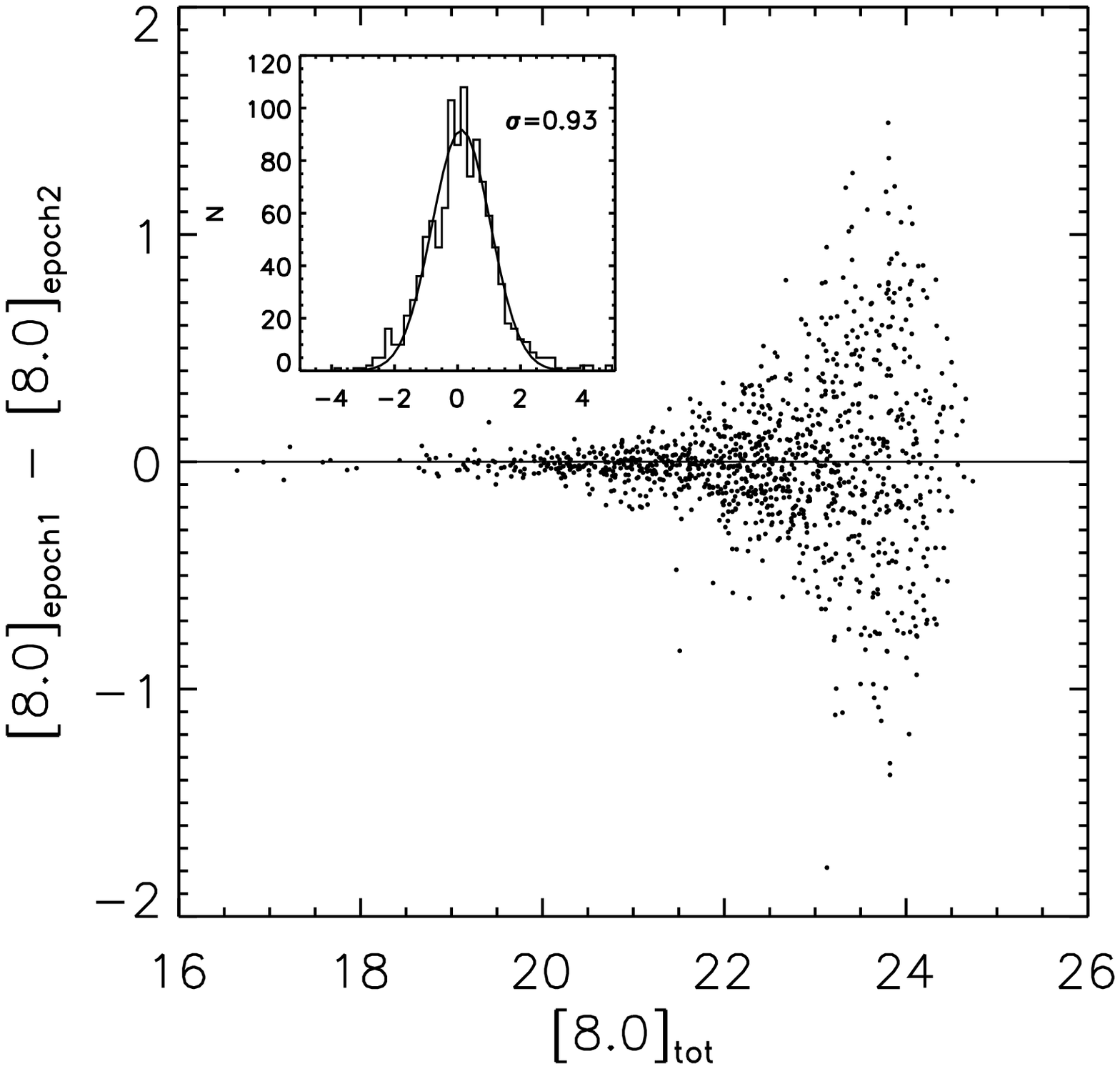}
\caption{\small Comparison between IRAC observations from epoch 1 and
epoch 2 for $K_s$-band detected sources in the overlap region between
the 2 epochs.  The large panels show a good correspondence between the
2 independent photometric measurements, with a slight zero-point drift
of 0.03 mag in the 3.6 $\mu$m band.  The inset panels shows the
distribution of $(f_{\rm epoch1} - f_{\rm epoch2}) / \sqrt{err_{\rm
epoch1}^2 + err_{\rm epoch2}^2}$, where a minimum relative error of
3\% was assumed to account for relative zero-point uncertainties over
the field.  The standard deviation of the distribution is of order
unity, meaning that estimated flux errors account well for the
empirically determined uncertainties.
\label {epochs.fig}
}
\end {figure*}

Uncertainties in the measured fluxes in the WFI/IRAC/MIPS bands have a
contribution from the background rms and from the residual
contamination of the subtracted neighbors (Labb\'{e} et al., in
preparation).  The former was calculated using the empty aperture
method described in \S\ref{bg_limdepth.sec}.  The latter was derived
by scaling the normalized convolved $K_s$-band image of each
neighboring source with the formal 1$\sigma$ error in its fitted
WFI/IRAC/MIPS flux as computed from the covariance matrix produced by
the least-squares minimization.  Subsequently, we performed aperture
photometry on this image of residual neighbor contamination.  This
gives the confusion error corresponding to the source flux measured
within an aperture of the same size.  The two contributions were added
in quadrature to obtain the total error budget.  In the 3.6 $\mu$m
band, where confusion is most severe, the confusion term is typically
5 times smaller than the background noise term.  For 6\% of the
sources in our catalog, the confusion term dominates the total error
budget.

Here, we follow an empirical approach to validate the size of the
uncertainties in the IRAC photometry.  We exploit the overlap region
between the 2 independent observation epochs of the CDFS with the IRAC
instrument.  The position angle was rotated over 180 degrees, causing
the PSF to have a different orientation with respect to the positions
of neighboring sources.  In Fig.\ \ref{epochs.fig}, we show the
difference between the IRAC magnitude measured during epoch 1 and
epoch 2.  The rms ranges from 5\% in the 4.5 $\mu$m band to 10\% in
the 8.0 $\mu$m band for sources with an AB magnitude brighter than 22.
The largest systematic offset was measured for the 3.6 $\mu$m band,
where a zero-point drift of 0.03 mag was measured between the 2
epochs.  In the inset panels the distribution of $(f_{\rm epoch1} -
f_{\rm epoch2}) / \sqrt{err_{\rm epoch1}^2 + err_{\rm epoch2}^2}$ is
plotted.  The distribution is well described by a gaussian.  For well
estimated errors the expected standard deviation of the distribution
is 1.  We adopted a minimum relative uncertainty in the flux of 3\% to
account for zero-point variations over the field.  This is
particularly relevant for the 3.6 $\mu$m and 4.5 $\mu$m band, where
the sources are detected with a high signal-to-noise.  The standard
deviation of $(f_{\rm epoch1} - f_{\rm epoch2}) / \sqrt{err_{\rm
epoch1}^2 + err_{\rm epoch2}^2}$ in these bands is smaller than 1.
Adopting a more conservative minimum relative uncertainty would only
decrease this value, suggesting that zero-point variations within the
field are limited to the few percent level.  In the less sensitive 5.8
$\mu$m and 8.0 $\mu$m bands, where the minimum relative uncertainty is
not reached, we find a distribution of $(f_{\rm epoch1} - f_{\rm
epoch2}) /
\sqrt{err_{\rm epoch1}^2 + err_{\rm epoch2}^2}$ with a standard deviation of
nearly unity, confirming empirically the validity of our estimated
uncertainties.

\section {Redshifts}
\label{redshifts.sec}

\subsection {Spectroscopic redshifts}
\label{specz.sec}

\begin{deluxetable}{lcc}
\tablecolumns{2}
\tablewidth{0pc}
\tablecaption{Spectroscopic redshifts for $K_s$-band detected objects \label{spec.tab}
}
\tablehead{
\colhead{Survey} & \colhead{High quality flags} & \colhead{Number\tablenotemark{a}}
}
\startdata
FORS2 (v3.0, Vanzella et al. 2008)    & A     & 357 \\
K20 (Mignoli et al. 2005)             & 1     & 265 \\
VVDS (v1.0, Le F\`{e}vre et al. 2004) & 4,3   & 251 \\
CXO (Szokoly et al. 2004)             & 3,2,1 & 89 \\
Norman et al. 2002                    & all   & 1 \\
Croom et al. 2001                     & all   & 21 \\
van der Wel et al. 2004               & all   & 24 \\
Cristiani et al. 2000                 & all   & 3 \\
Strolger et al. 2004                  & all   & 5 \\
Daddi et al. 2004                     & all   & 7 \\
IMAGES (Ravikumar et al. 2006)        & 1     & 100 \\
LCIRS (Doherty et al. 2005)           & 3     & 3 \\
Wuyts et al. (in prep.)               & all   & 7 \\
Kriek et al. (2007)                   & all   & 2 \\
Roche et al. (2006)                   & all   & 4 \\
Huang et al. (in prep.)               & all   & 9 \\
VIMOS (v1.0, Popesso et al. 2008)     & A     & 329
\enddata
\tablenotetext{a}{The numbers are non-redundant.  For objects targeted during multiple surveys, the redshift with the highest quality flag was adopted.}
\end{deluxetable}

The CDFS-GOODS area has been targeted intensively by various
spectroscopic surveys, listed in Table\ \ref{spec.tab}.  The combined
sample of spectroscopic redshifts forms a heterogeneous family of
objects, with selection criteria varying from pure $I$-band (VVDS, Le
F\`{e}vre et al. 2004), $K_s$-band (Mignoli et al. 2005) or X-ray
(Szokoly et al. 2004) flux limits to various color criteria (e.g.,
Doherty et al. 2005; Wuyts et al. in preparation).  It is therefore
impossible to build a complete $K_s$-band selected spectroscopic
sample from the data at hand.  Rather, we aim to provide a list of
trustworthy spectroscopic redshifts that are reliably cross-identified
with a $K_s$-band detection in our catalog.  To do so, we apply a
conservative quality cut based on the quality flags that come with
each of the spectroscopic catalogs, and assign the redshift to the
nearest $K_s$-band selected object within a radius of $1\farcs 2$.
The quality flags and number of sources included in our reliable list
of cross-correlated spectroscopic redshifts are summarized in Table\
\ref{spec.tab}.  We mark these sources with a ``zsp\_qual'' flag of 1 in
our catalog.  For completeness, other spectroscopic redshifts for
$K_s$-band detected objects are also listed in our catalog, marked
with a ``zsp\_qual'' flag lower than 1 (zsp\_qual=0.5 for FORS2
quality flag B and VIMOS quality flag B, and zsp\_qual=0.1 for all
others), together with the original quality flag from the respective
survey.  We proceed to use only the 1477 spectroscopic redshifts with
zsp\_qual = 1.

\subsection {Photometric redshifts}
\label{photz.sec}
Together with the observed photometry, we release a list of
photometric redshifts computed with the new photometric redshift code
EAZY.  A full description of the algorithm and template sets will be
presented by Brammer et al. (in preparation).  Briefly, the program
fits a non-negative superposition of SED templates to the
$U_{38}$-to-8 $\mu$m photometry, using a template error function that
effectively downweights the rest-frame UV and rest-frame NIR of the
templates in the fit.  Next, a redshift probability distribution
$p(z|C,m_0)$ is constructed for each galaxy with observed colors C,
using the $K_s$-band magnitude $m_0$ as a prior.  Finally, the value
$z_{\rm mp}$ of the redshift marginalized over the total probability
distribution,
\begin{equation}
z_{\rm mp} = \frac{\int z\ p(z|C,m_0)\ dz} {\int p(z|C,m_0)\ dz},
\end{equation}
was adopted as best estimate of the galaxy's redshift.  1$\sigma$
confidence limits were computed by integrating the redshift
probability distribution from the edges till the integrated
probability equals $0.317/2$.

The template set constructed by Brammer et al. (in preparation) is
based on P\'{E}GASE models (Fioc \& Rocca-Volmerange 1997).  From a
library of $\sim 3000$ templates described by G06a, Brammer et al. (in
preparation) followed the ``non-negative matrix factorization''
algorithm of Blanton \& Roweis (2007) in order to derive a set of 5
principal component templates that span the colors of galaxies in the
semi-analytic model by De Lucia \& Blaizot (2007).  One dusty template
with an age of 50 Myr and $A_V=2.75$ was added to account for the
existence of dustier (and thus redder) galaxies than present in the
semi-analytic model mock catalog.  It is this set of 6 spectral
templates that we fed to EAZY.

\begin {figure} [htbp]
\centering
\plotone{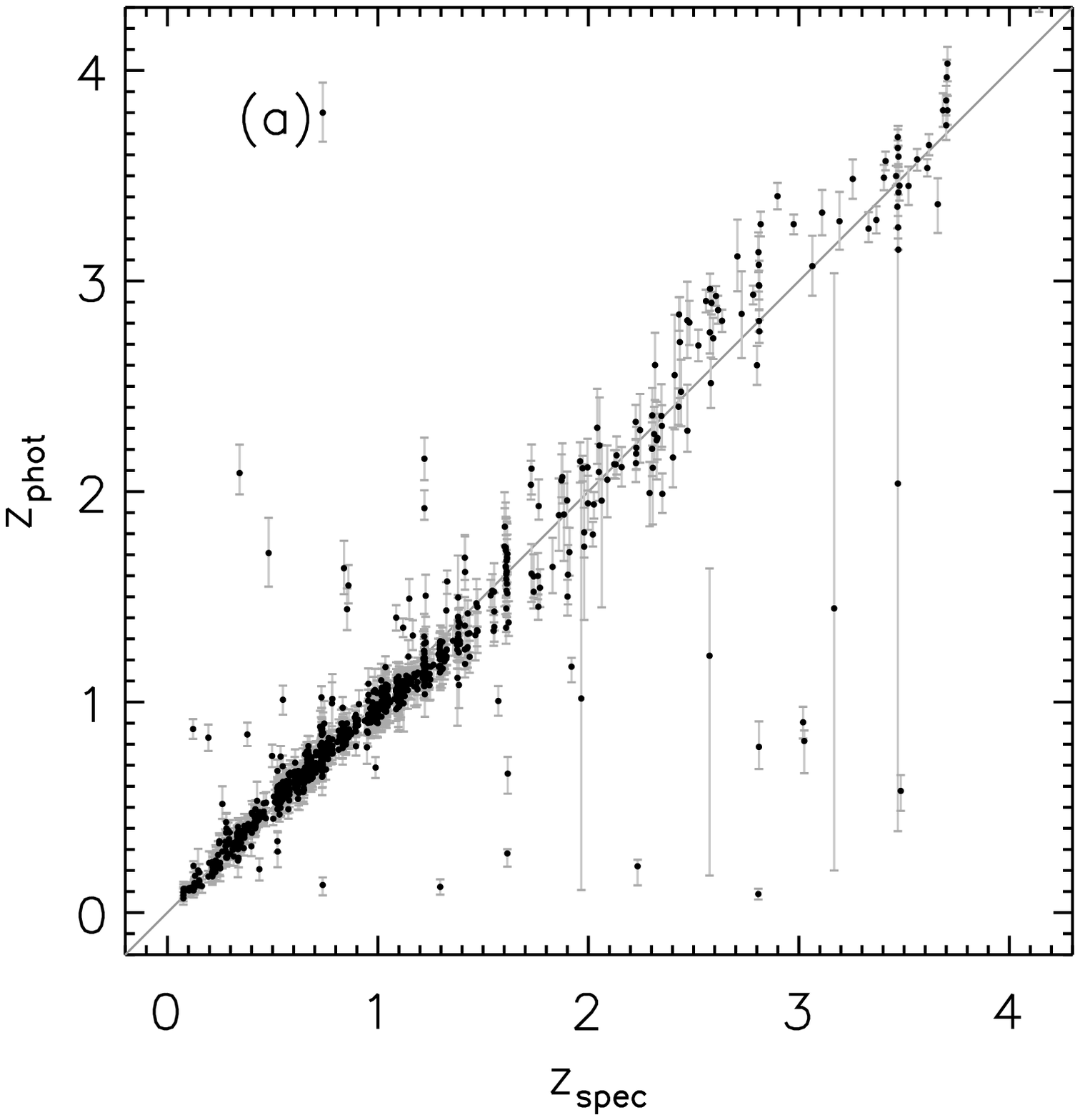}
\vspace{0.1in}
\plotone{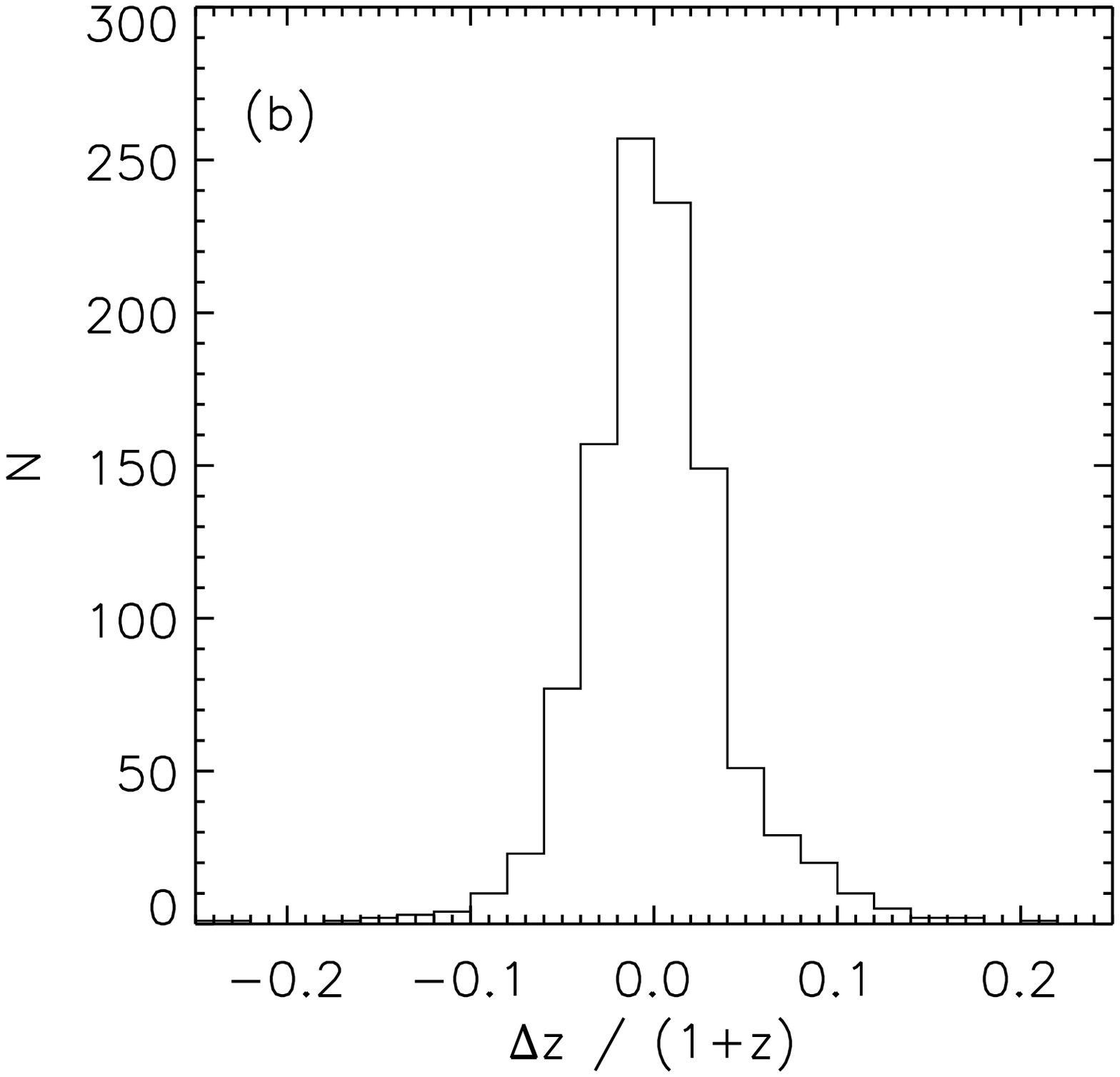}
\caption{\small Comparison between photometric and spectroscopic
redshifts for 1067 $K_s$-band detected sources with reliable $z_{\rm spec}$
identification and coverage in all wavelength bands.  (a) A direct
comparison with 1$\sigma$ confidence intervals derived from the
redshift probability distribution.  (b) The distribution of $\Delta z
/ (1+z)$.  2.5\% of the sources fall outside the plotted range.
\label {zspzph.fig}
}
\end {figure}

We quantify the accuracy of the photometric redshifts $z_{\rm phot}$ by
a comparison to the spectroscopic redshifts with zsp\_qual = 1.  Fig.\
\ref{zspzph.fig}(a) shows the correspondence between $z_{\rm phot}$ and
$z_{\rm spec}$ for all 1067 sources with $K_{s,{\rm AB}}^{\rm
tot}<24.3$ that are covered by all bands and for which a reliable
spectroscopic redshift is available.  The error bars indicate the
1$\sigma$ confidence intervals as derived from the redshift
probability distribution.  We find that the spectroscopic redshift
lies within the formal 1$\sigma$ confidence interval for 66\% of the
sources, confirming that the uncertainties on $z_{\rm phot}$ have reliably
been established.  Fig.\
\ref{zspzph.fig}(b) presents the distribution of $\Delta z / (1+z)$, with $\Delta z = z_{\rm phot} - z_{\rm spec}$,
which is commonly used to determine the accuracy of photometric
redshifts.  We find a median $\Delta z / (1+z)$ of -0.001 and a
normalized median absolute deviation (equal to the rms for a Gaussian
distribution) of $\sigma_{\rm NMAD}=0.032$.  3\% of the objects with
spectroscopic redshift have $|\Delta z| / (1+z) > 5\sigma_{\rm NMAD}$.

We investigated the quality of $z_{\rm phot}$ estimates as a function of
redshift and total $K_s$-band magnitude.  Variations in the median
$\Delta z / (1+z)$ as function of redshift are small, reaching a
maximum of 0.05 in the $2.5<z<3$ interval.  No obvious dependence on
$K_{s,{\rm AB}}^{\rm tot}$ is observed.  The scatter in $\Delta z / (1+z)$ is
roughly a factor 2.4 larger for the $z>1.5$ regime
($\sigma_{\rm NMAD}=0.071$) than for the $z<1.5$ ($\sigma_{\rm NMAD}=0.030$)
regime.

Considering the 98 spectroscopically confirmed sources with a
cross-identification within $2\farcs 0$ in the 1Ms X-ray catalog by
Giacconi et al. (2002), we find a scatter of $\sigma_{\rm NMAD}=0.043$.
It is reassuring that despite the lack of AGN spectrum in our template
set, the netto performance of our photometric redshift code for AGN
candidates remains good.  We do note however that, independent of
redshift, the fraction of catastrophic outliers ($|\Delta z| / (1+z) >
5\sigma_{\rm NMAD}$) is 3 times larger for the AGN candidates than for
the total sample of spectroscopically confirmed sources.

\section {Catalog parameters}
\label{param.sec}
Here we describe the entries of our $K_s$-band selected FIREWORKS
catalog of the GOODS-CDFS.  The format is similar to the FIRES
catalogs of the HDFS (L03) and \1054 (FS06), making a straightforward
combination of all three fields possible for the user.  The catalog
can be obtained from the FIREWORKS
homepage\footnote[1]{http://www.strw.leidenuniv.nl/fireworks}.

\begin {itemize}
\item ID-- Unique identification number
\item x, y-- Pixel position of the object, based on the $K_s$-band detection map.  The pixel scale is $0\farcs 15$ pixel$^{-1}$.
\item RA, DEC-- Right ascension and declination coordinates for equinox J2000.0.
\item ${\it [band]}$\_colf-- Flux in microjanskys measured within the color aperture (\S\ \ref{opt_NIR_phot.sec}).  The bandpasses are $U_{38}$,$B_{435}$,$B$,$V$,$V_{606}$,$R$,$i_{775}$,$I$,$z_{850}$,$J$,$H$,$K_s$, $[3.6\mu$m$]$, $[4.5\mu$m$]$, $[5.8\mu$m$]$, and $[8.0\mu$m$]$.
\item ${\it [band]}$\_colfe-- Uncertainty in the ${\it [band]}$\_colf flux measurement, derived from the noise analysis (\S\ \ref{bg_limdepth.sec}).  The units are microjanskys.
\item $K_s$\_totf-- Total $K_s$-band flux in microjanskys, measured within the total aperture and scaled by the aperture correction (\S\ \ref{opt_NIR_phot.sec}).  Approximate total fluxes in other bandpasses can be computed by ${\it [band]}\_totf = {\it [band]}\_colf \times (K_s\_totf / K_s\_colf)$, provided there are no large morphological k-corrections.
\item $K_s$\_totfe-- Uncertainty associated with $K_s$\_totf, also in microjanskys.
\item $[24\mu$m$]$\_totf-- Total MIPS 24 $\mu$m-band flux in microjanskys, measured within a 6'' diameter circular aperture and then aperture corrected (\S\ \ref{MIR_phot.sec}).
\item ${\it [band]}$w-- Effective weight in the bandpass ${\it [band]}$, normalized to the median effective weight of all sources in that band.  We recommend applying a conservative weight criterion ${\it [band]}$w $> 0.3$ to construct samples with robust photometry.
\item ap\_col-- Aperture diameter in arcsec within which ${\it [band]}$\_colf was measured.  In cases where the color aperture was the isophotal aperture defined by the surface brightness threshold of $\mu (K_{s,{\rm AB}}) = 24.6$ mag arcsec$^{-2}$, ap\_col is the diameter in arcsec of a circular aperture with equal area.
\item ap\_tot-- Aperture diameter in arcsec used for measuring the total $K_s$-band flux.  When the isophotal or SExtractor's ``MAG\_AUTO'' aperture was used, this entry contains the equivalent circularized diameter corresponding to that aperture.
\item f\_deblend1-- Flag equal to 1 when the source was deblended somewhere in the process (SExtractor's ``blend'').
\item f\_deblend2-- Flag equal to 1 when the photometry is affected by a neighboring source (SExtractor's ``bias'').
\item Kr50-- Half light radius in arcsec, measured on the $K_s$-band image (SExtractor's flux\_radius scaled to arcsec).
\item Keps-- Ellipticity of the isophotal area, measured on the $K_s$-band image.
\item Kposang-- Position angle of the isophotal area, measured on the $K_s$-band image.
\item zph\_best-- Best estimate of the photometric redshift (\S\ \ref{photz.sec}).
\item zph\_low, zph\_high-- Lower and upper edge of the 68\% confidence interval around zph\_best.
\item zsp-- Spectroscopic redshift (set to -99 when no spectroscopic information is available).
\item zsp\_qual-- Quality flag from 0 to 1 assigned to the spectroscopic redshift.  Only zsp\_qual=1 entries are considered reliable.
\item zsp\_source-- Spectroscopic survey from which zsp was taken (Tab.\ \ref{spec.tab}).
\item zsp\_qual\_orig-- Original quality flag for zsp from the respective spectroscopic survey.
\item XID-- Identification number from the 1Ms X-ray catalog by Giacconi et al. (2002), set to -99 when no cross-identification within 2'' was found.  Note that we accounted for the $\sim 1\farcs 3$ systematic offset in the Giacconi et al. (2002) X-ray centroids, as pointed out before by Roche et al. (2003).
\end {itemize}

\section {Comparison to the GOODS-MUSIC catalog}
\label{comparison.sec}

\subsection {Differences in data and strategy}
Recently, G06a presented a multicolor catalog for
the GOODS-CDFS field, referred to as the GOODS-MUSIC catalog.  The
clustering evolution of distant red galaxies was quantified based on
this sample (Grazian et al. 2006b), as was the contribution of various
color-selected samples of distant galaxies to the stellar mass density
(Grazian et al. 2007).  Despite the overlap in public data used to
compile the GOODS-MUSIC and our catalog, there are a number of marked
differences.

First, our catalog is purely $K_s$-band selected.  Since the
$K_s$-band magnitude is a better proxy for stellar mass than the
optical magnitude, this makes it ideally suited to extract
mass-limited samples from.  The GOODS-MUSIC sample on the other hand
is to first order $z_{850}$-band selected (at the ACS resolution), with an
addition of the remaining $K_s$-band sources that are detected in a
map with masked $z_{850}$-band detections.  Although valuable in its own
respect, this makes it less trivial to understand the completeness of
the sample.

Second, we based our catalog on the ESO/GOODS data release v1.5,
consisting of 3 extra ISAAC pointings in $J$ and $K_s$, and 7 more in $H$
with respect to the v1.0 release used by G06a.

Third, apart from the $U_{38}$ photometry, we also added photometry
based on images in the other WFI passbands: $BVRI$.  Despite the
shallower depth than the ACS imaging (see Table\ \ref{dataset.tab}),
this addition is useful for, e.g., estimating photometric redshifts,
since the WFI passbands are nicely centered in between the ACS
passbands.  Consequently, the spectral energy distributions of sources
in the catalog are better sampled.

Finally, we include MIPS 24 $\mu$m measurements, enabling us to
constrain the total IR luminosity of the $K_s$-band selected
galaxies.  Before doing so, we compare the photometry in common
between both catalogs, and the photometric redshifts derived from it.

\subsection{Comparing photometry}
\label{MUSICphot.sec}

\begin {figure*} [htbp]
\centering
\epsscale{1}
\plotone{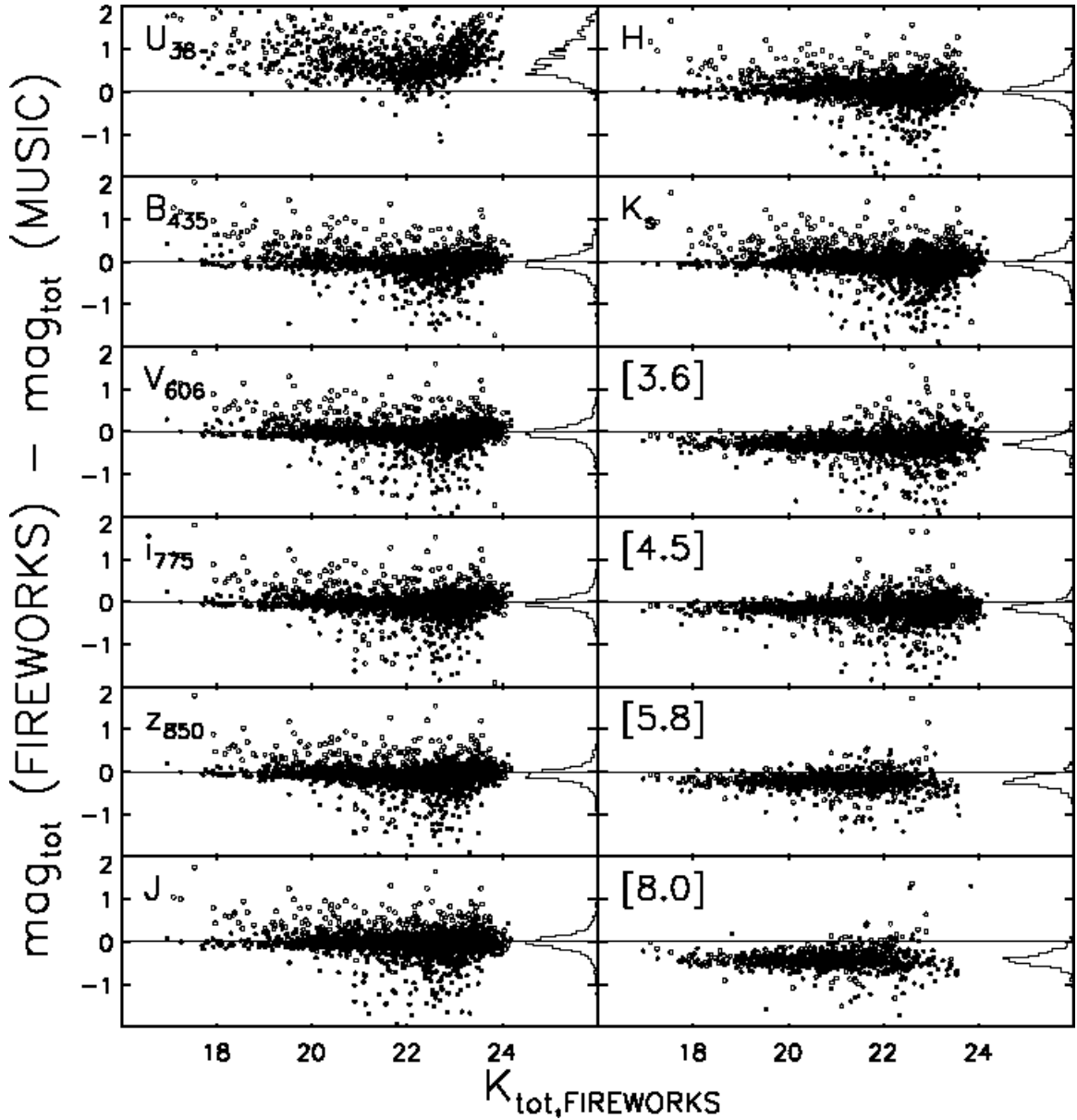}
\caption{\small A direct comparison of total magnitudes for sources
with $S/N > 10$ in the $U_{38}$-to-8.0 $\mu$m bandpasses in common
between GOODS-MUSIC and our FIREWORKS catalog.  Sources that are
blended in the $K_s$-band image are plotted as empty symbols.  On the
right side of each panel, a histogram shows the distribution of
offsets.  Our $U_{38}$ photometry, based on the re-reduced WFI image
by the GaBoDS consortium, is systematically fainter by 0.2 mag.  We
find an overall good correspondence in the ACS optical and ISAAC NIR
bands, with offsets of roughly 6\% due to aperture corrections.  In
the IRAC bands, photometric offsets amount up to 0.4 mag.
\label {comp_phot.fig}  
}
\end {figure*}

We cross-correlated the two catalogs using a search radius of $1\farcs
2$ and in Fig.\ \ref{comp_phot.fig} present a comparison of the
$U_{38}$-to-8.0 $\mu$m photometry for the bands in common between
them.  Differences between the total magnitudes are plotted for
objects with $S/N > 10$ in the $K_s$-band and the band under
consideration.  Objects that are marked by SExtractor as blended in
the $K_s$-band, are indicated with empty symbols.  

Our $U_{38}$ magnitudes for non-blended sources are fainter than the
GOODS-MUSIC magnitudes by 0.2 mag in the median, with a scatter of 0.4
mag.  This offset can be attributed to our use of the re-reduced WFI
data for which the GaBoDS consortium also supplied an improved
zeropoint determination.

The overall correspondence in the $B$-to-$K_s$ bands is good, and
offsets can be well understood from the differences in the applied
photometric method.  We measure a typical median offset for
non-blended sources in the optical and NIR bands of $mag_{\rm tot,FIREWORKS} -
mag_{\rm tot,MUSIC} = -0.06$, and a scatter of $\sigma_{\rm NMAD} < 0.2$.
G06a based their total magnitudes on SExtractor's ``MAG\_AUTO''
parameter for the $z_{850}$-band detections and on the ``MAG\_BEST''
for the remaining $K_s$-band detections that were not detected in the
$z_{850}$-band.  G06a did not apply an aperture correction based on the
stellar growth curve to correct for the flux lost because it fell
outside the ``MAG\_AUTO'' or ``MAG\_BEST'' aperture.  The lack of
aperture correction explains at least part of the systematic offset.
Sources marked as blended in our $K_s$-band detection map typically
are brighter by 0.2 - 0.4 mag in the MUSIC catalog.  This can be
explained by the contamination from neighboring sources within the
``MAG\_AUTO'' aperture, which we avoid by using the isophotal aperture
in combination with an aperture correction for blended sources.
However, our measure of the total magnitude in the presence of
crowding is also not perfect in that flux from the source that is
projected on top of a blended source may be lost by our technique.

For the IRAC photometry, the discrepancies are larger, ranging from
0.16 mag in the 4.5 $\mu$m band to 0.42 mag at 8.0 $\mu$m.  The
observed offsets in IRAC photometry can largely be attributed to the
use of an early version of the IRAC PSF by G06a and a bug in the
normalization of the smoothing kernel for the IRAC data by G06a
(Fontana \& Grazian, private communication).  After removing these two
effects, our measurements of the total IRAC magnitudes are still
brighter by 0.06 to 0.1 mag.  This offset is similar to that for the
optical and NIR bands, and can be attributed to light missed by the
``MAG\_BEST'' magnitude, as G06a already caution.  We stress that the
offset in total IRAC magnitudes not only affects its derived
properties such as stellar mass, but also the optical-to-MIR and
NIR-to-MIR colors.  For example, our $z_{850} - [3.6\mu$m$]$, $z_{850}
- [4.5\mu$m$]$, $z_{850} - [5.8\mu$m$]$, and $z_{850} - [8.0\mu$m$]$
colors are redder than the GOODS-MUSIC colors by 0.23, 0.11, 0.17, and
0.37 mag in the median respectively.  Similarly, our $K_s -
[3.6\mu$m$]$, $K_s - [4.5\mu$m$]$, $K_s - [5.8\mu$m$]$, and $K_s -
[8.0\mu$m$]$ colors are redder in the median by 0.29, 0.16, 0.22, and
0.42 respectively.  The scatter in the color differences with respect
to GOODS-MUSIC typically amounts to 1.5 times the size of the median
offset.

\subsection{Comparing photometric redshifts}
Finally, we compare the photometric redshifts presented in
\S\ref{photz.sec} with those derived by G06a.  The
numbers quoted in \S\ref{photz.sec} and by G06a cannot directly be
compared since new spectroscopic redshifts were added, and objects
that showed evidence for the presence of an AGN in their optical
spectrum were rejected from the GOODS-MUSIC photometric redshift
analysis.  Nevertheless, when comparing the performance of the
$z_{\rm phot}$ estimates for a set of 659 non-AGN with reliable $z_{\rm spec}$
and coverage in all bands in both catalogs, we find a scatter in
$\Delta z / (1+z)$ for GOODS-MUSIC ($\sigma_{\rm NMAD}=0.039$) that is
similar to that for our best estimates ($\sigma_{\rm NMAD}=0.033$).  The
median $\Delta z / (1+z)$ is -0.009 and -0.005 for the GOODS-MUSIC and
FIREWORKS $z_{\rm phot}$ estimates respectively.

Comparing the $z_{\rm phot}$ estimates for all objects that are well
exposed in all bands in both catalogs, irrespective of a spectroscopic
confirmation, we find a median $\frac{z_{\rm phot,MUSIC} - z_{\rm
phot,FIREWORKS}}{1 + z_{\rm phot,FIREWORKS}}$ of 0 and a scatter
$\sigma_{\rm NMAD} = 0.052$. The largest systematic offsets occur in
the $1.5<z<2$ and $2.5<z<3$ intervals where median($\frac{z_{\rm
phot,MUSIC} - z_{\rm phot,FIREWORKS}}{1 + z_{\rm phot,FIREWORKS}}$)
$\sim$ 0.04 and -0.04 respectively.  We find no systematic dependence
on $K_{s,{\rm AB}}^{\rm tot}$.

We conclude that the photometric differences between both catalogs can
be understood from the applied method.  Our $z_{\rm phot}$ estimates are
in excellent agreement with the available spectroscopic samples and
generally show good agreement with the $z_{\rm phot}$ estimates by G06a.
We now proceed to exploit our catalog to analyze the colors and total IR
energy output of distant galaxies.

\section {Total IR properties of distant $K_s$-selected galaxies}
\label{science.sec}

With the catalog at hand, we aim to answer the following simple
questions: Which $K_s$-selected ($S/N_{K_s} > 5$, $K_{s,{\rm AB}}^{\rm tot}
\lesssim 24.3$) galaxies at $1.5<z<2.5$ have the brightest total IR luminosities ($L_{\rm IR} \equiv
L(8-1000\ \mu$m$)$), and which contribute most to the integrated total
IR luminosity?  Specifically we will address whether the total IR
luminosity is dominated by red or blue galaxies, with the color
defined in the rest-frame UV, optical or NIR wavelength regime.  We
restrict our analysis to the $1.5<z<2.5$ interval, since at those
redshifts the observed 24 $\mu$m probes the rest-frame mid-IR ($7\
\mu$m$\lesssim \lambda_{\rm rest} \lesssim 10\ \mu$m), which broadly
correlates with the total IR luminosity (e.g., Spinoglio et al. 1995;
Chary \& Elbaz 2001; Dale \& Helou 2002).

\subsection {Observed 24 $\mu$m flux as function of observed colors}
\label{science_obs.sec}
We approach the questions raised above by first studying the
correlation between purely observational properties: the 24 $\mu$m
flux as proxy for IR luminosity and the observed $B_{435}-V_{606}$, $J-K_s$, and
$K_s - [4.5\mu$m$]$ colors as proxy for the rest-frame UV, optical and
optical-to-NIR color respectively.  Unless the redshift dependence of
the conversion from 24 $\mu$m to total IR luminosity and of the
conversion from observed to rest-frame colors are conspiring, any
trend in the directly observable properties should be a signpost for
correlations in the rest-frame properties, whose derivation involves
significant systematic uncertainties.

\begin {figure} [htbp]
\centering
\plotone{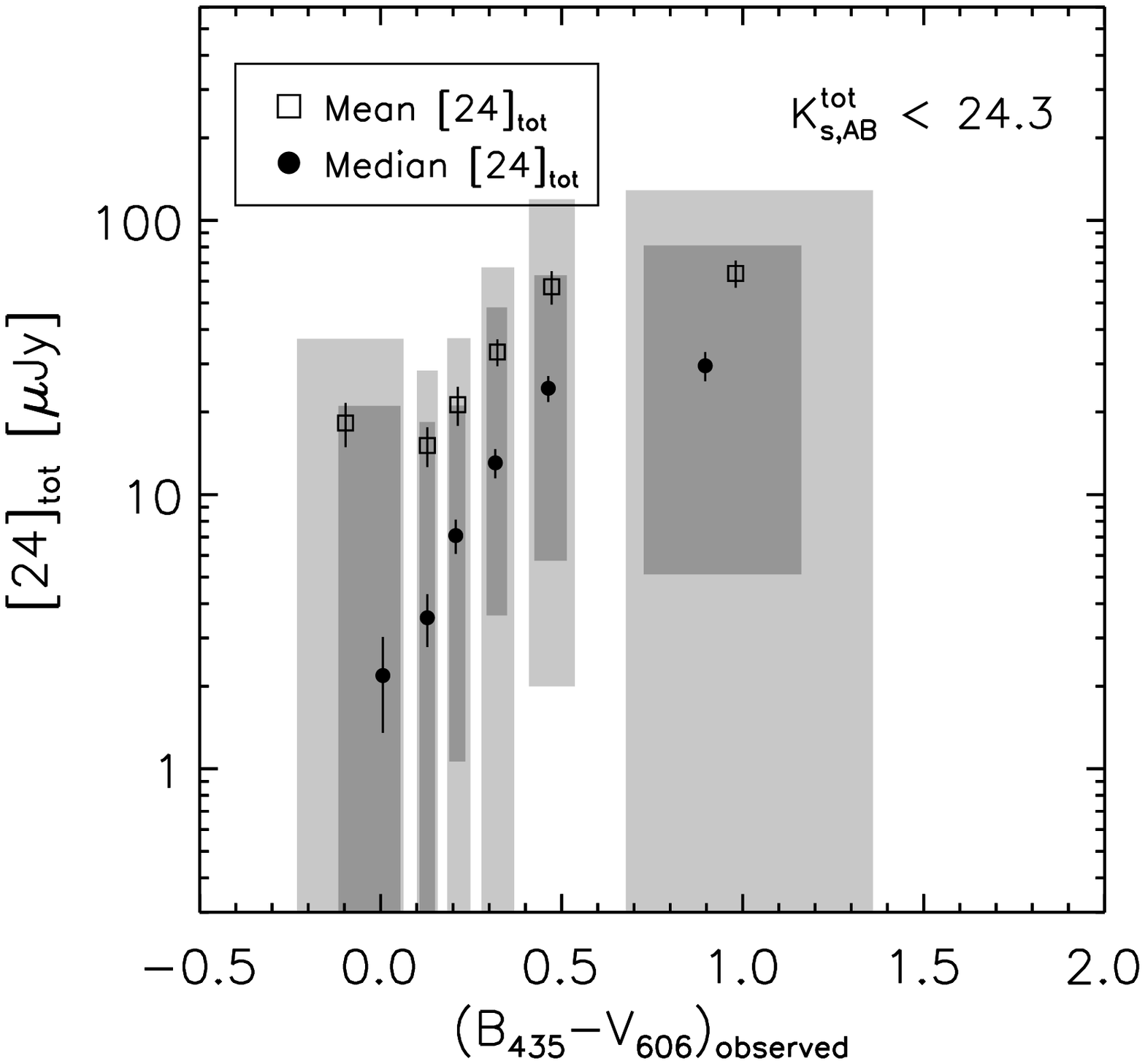}
\plotone{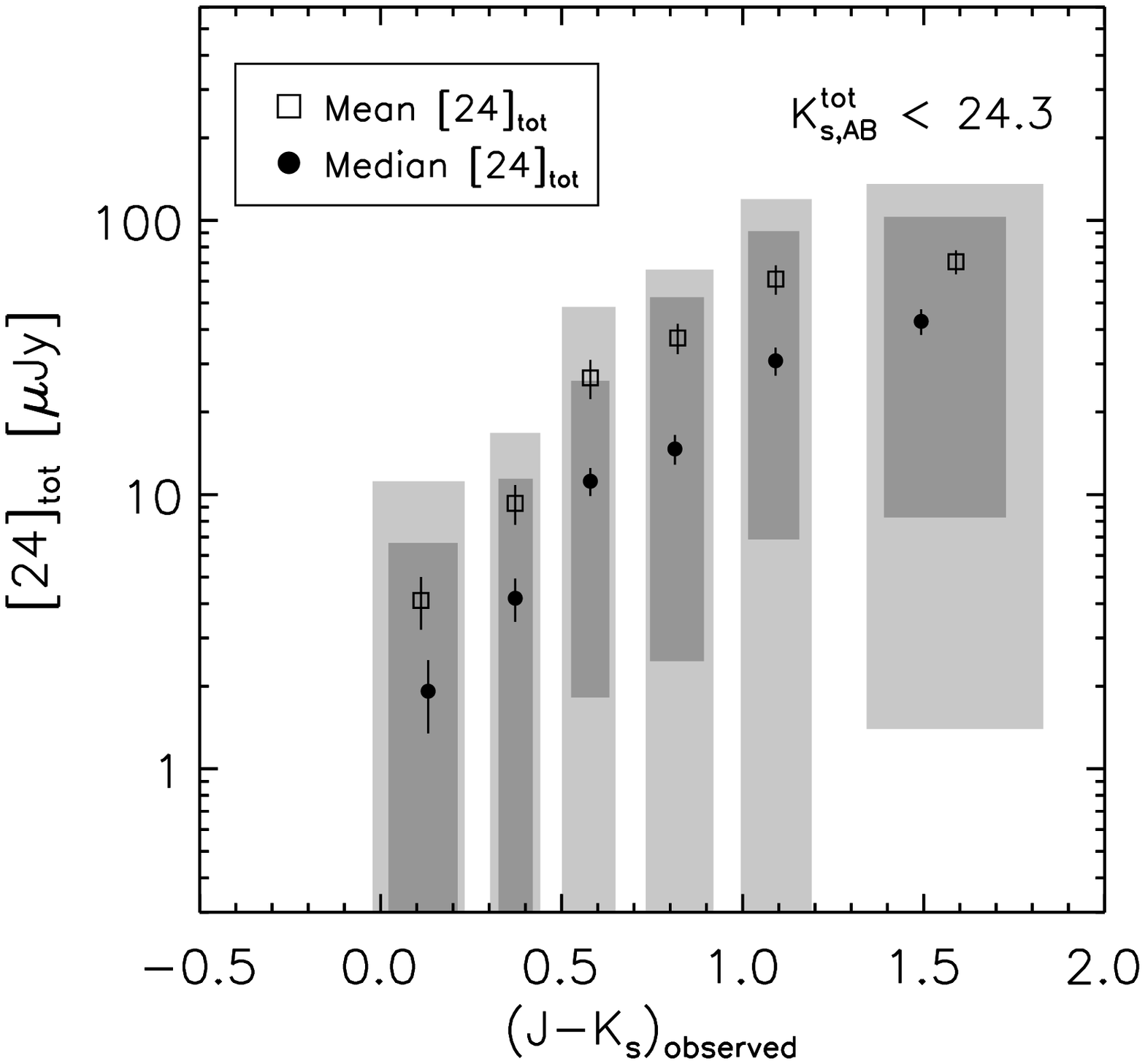}
\plotone{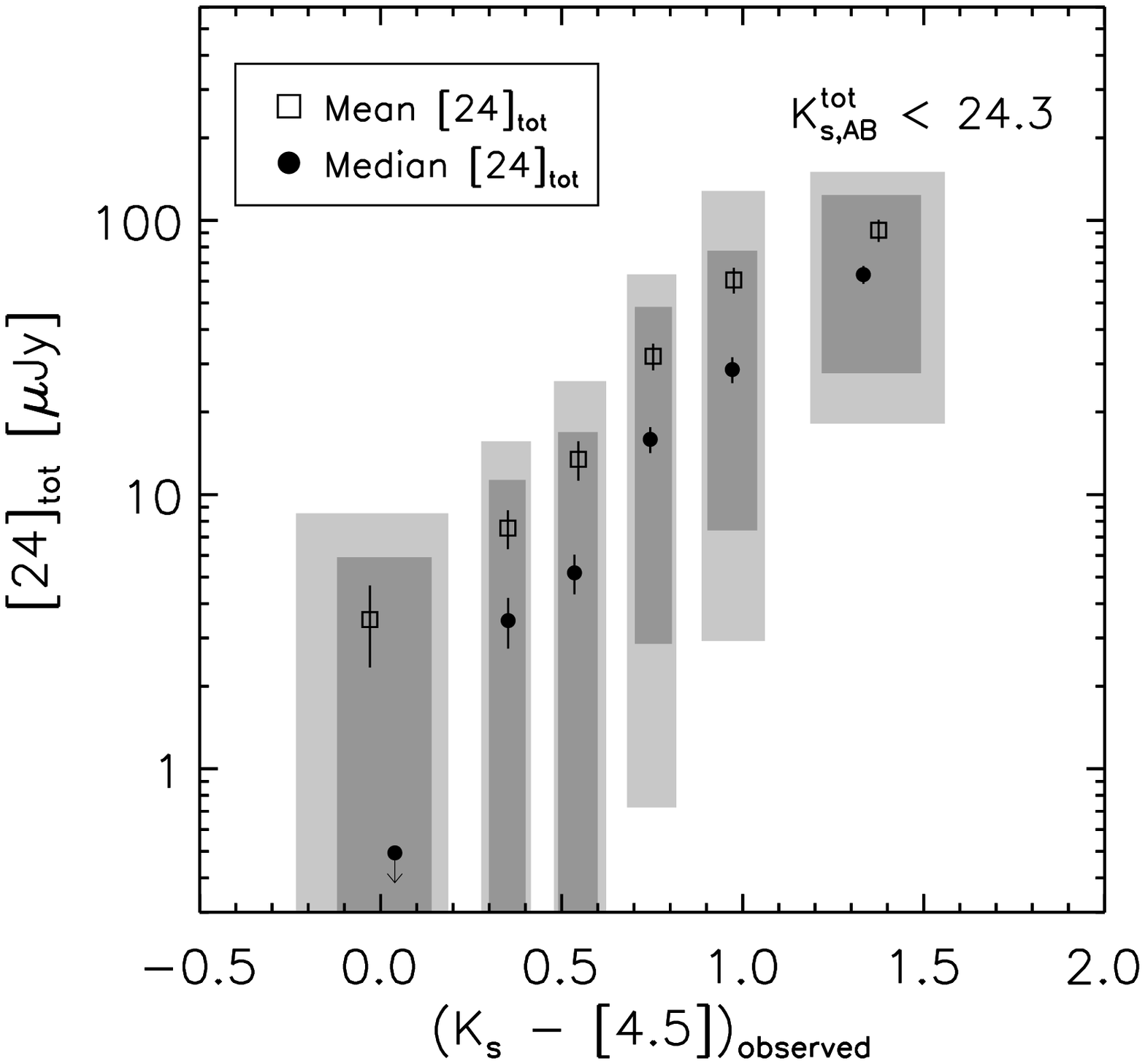}
\caption{\small Stacked 24 $\mu$m flux densities as function of
observed $B_{435}-V_{606}$, $J-K_s$, and $K_s - [4.5\mu$m$]$ color for galaxies at
$1.5<z<2.5$ with $S/N > 5$ in the $K_s$-band (corresponding to
$K_{s,{\rm AB}}^{\rm tot} \lesssim 24.3$).  Filled circles represent the median
for each equal-number bin.  Empty boxes represent the mean flux.
Light-grey and dark-grey polygons indicate respectively the central
68\% and 50\% of the distribution within each bin.  We find a trend of
increasing $[24\mu$m$]_{\rm tot}$ with redder observed-frame color that
gets progressively stronger as we consider colors measured at longer
wavelengths.
\label {MIPSobs.fig}  
}
\end {figure}

The total number of $K_s$-selected galaxies at $1.5<z<2.5$ with
$K^{\rm tot}_{s,{\rm AB}} < 24.3$ is 961.  45\% of these are individually
detected at the $3\sigma$ level ($[24\mu$m$]_{\rm tot} > 11\
\mu$Jy).  Since we aim to investigate what color galaxies dominate the IR
emission, we divide our galaxies in bins of similar color.  Each bin
contains an equal number of sources (160).  Consequently, the bin
widths are not equal.  To start, we leave the origin of the 24 $\mu$m
emission (dust heated by AGN or star formation) as an open question.
We note however that when excluding X-ray detected sources each bin would
contain 154 objects, and applying such a selection would not affect
the results of our analysis.  We perform the stacking per color bin in
two ways: by averaging the 24 $\mu$m flux measurements of all sources
in the bin (some of which will be negative due to noise), and by
taking the median of the 24 $\mu$m flux measurements of all sources in
the bin.  The mean has a contribution from all the galaxies in the
color bin, and hence directly measures the contribution to the
integrated 24 $\mu$m emission.  The median has the advantage that it
is more robust against a small number of bright sources in the color
bin, and is therefore lower.

In Fig.\ \ref{MIPSobs.fig}, the stacked 24 $\mu$m flux densities are
plotted versus the observed $B_{435}-V_{606}$, $J-K_s$, and $K_s - [4.5\mu$m$]$
color.  The error bars on the mean flux measurement indicate the
errors in the mean ($\sigma([24]_{\rm tot}) /
\sqrt{N}$), whereas the error bars on the median flux measurement are
computed as $\sigma_{\rm NMAD}([24]_{\rm tot}) /
\sqrt{N}$.  Furthermore, the light grey and dark grey polygon show the
range containing respectively the central 68\% and 50\% of the binned
galaxies.  Each color bin contains galaxies with a large spread in 24
$\mu$m fluxes.  In most bins, at least 20\% of the galaxies have an
individual signal-to-noise ratio $S/N < 1$ at 24 $\mu$m.

Fig.\ \ref{MIPSobs.fig}(a) shows that the galaxies in the bluest
$B_{435}-V_{606}$ bins are the faintest 24 $\mu$m sources.  However,
the stacked $[24\mu$m$]_{\rm tot}$ flux is not uniformly increasing over
the whole observed optical color range.  Considering colors measured
at longer wavelengths, we find a highly significant increase in the
stacked $[24\mu$m$]_{\rm tot}$ flux over the entire $J-K_s$ (Fig.\
\ref{MIPSobs.fig}(b)) and $K_s - [4.5\mu$m$]_{\rm tot}$ (Fig.\
\ref{MIPSobs.fig}(c)) color range.  The trend is strongest in the
observed $K_s - [4.5\mu$m$]_{\rm tot}$ color, where we find an increase in
$[24\mu$m$]_{\rm tot}$ of a factor 26 in the mean and 2 orders of
magnitude in the median over a color range of $\sim 1.3$ mag.  Since
the bins contain an equal number of objects, it is trivial to see that
not only the reddest galaxies in $J-K_s$ and $K_s-[4.5\mu$m$]$ are
brightest at 24 $\mu$m, they also contribute the most to the total 24
$\mu$m emission integrated over all $K_s$-selected galaxies at
$1.5<z<2.5$.

\subsection {Total IR luminosity as function of rest-frame colors}
\label{science_mod.sec}

Although the trend of more 24 $\mu$m emission for galaxies with a
redder observed color is highly significant for $J-K_s$ and $K_s -
[4.5\mu$m$]$, it could still be contaminated or, alternatively, driven
by redshift dependencies within the $1.5<z<2.5$ redshift interval
under consideration.  Now, we will attempt to remove possible redshift
dependencies by converting both axes to a rest-frame equivalent.
Moreover, instead of converting the measured flux density at 24 $\mu$m
to a rest-frame flux density at 24 $\mu$m$ / (1+z)$, we use it as a
probe to determine the total IR luminosity $L_{\rm IR} \equiv L(8-1000\
\mu$m$)$.  Since this conversion assumes that the 24 $\mu$m emission
originates from dust heated by star formation, we further reject all
X-ray detections from our sample to rule out relatively unobscured AGN
candidates.  We caution that among the brightest 24 $\mu$m sources in
our sample, there may still be cases where hot dust emission from an
obscured AGN contributes significantly to the 24 $\mu$m emission and
causes an overestimate of $L_{\rm IR}$.  Detailed studies of such mid-IR
excess sources are presented by Daddi et al. (2007) and Papovich et
al. (2007).  Papovich et al. (2007) find that for sources with
$[24\mu$m$]_{\rm tot} >$ 250 $\mu$Jy the $L_{\rm IR}$ estimates derived from
the 24 $\mu$m flux density alone are too large by factors 2-10.  We
stress however that we study a deep $K_s$- and not 24 $\mu$m-selected
sample.  The fraction of sources with $[24\mu$m$]_{\rm tot} >$ 250 $\mu$Jy
per color bin never exceeds 7\%.  Therefore, at least the median
stacked properties are relatively robust against a possible
contamination by obscured AGN.

In the following, we first describe the derivation of rest-frame UV to
NIR colors.  Next, we explain the method to estimate the total IR
luminosity.  Finally, we repeat the stacking analysis using the
derived rest-frame properties.

\subsubsection {Stellar mass, UV slope, and rest-frame colors}
\label{beta_rfcol.sec}
For each of the galaxies in our sample, we modeled the spectral energy
distribution (SED) using the stellar population synthesis code by
Bruzual \& Charlot (2003).  We used an identical approach as Wuyts et
al. (2007), assuming a Salpeter IMF and solar metallicity, and fitting
three star formation histories: a single stellar population without
dust, an exponentially declining model with {\it e}-folding time of
300 Myr and allowed dust attenuation in the range $A_V = 0 - 4$, and a
constant star formation model with the same freedom in attenuation.
We subsequently scaled the stellar masses derived from the
best-fitting template to a Kroupa IMF by dividing the stellar masses
for a Salpeter IMF by a factor 1.6.  

We characterize the rest-frame UV part of each SED by fitting the
functional form $F_{\lambda} \sim
\lambda^{\beta}$ to the best-fitting template, using the rest-frame UV
bins defined by Calzetti, Kinney,\& Storchi-Bergmann (1994).  The
robustness of this technique is discussed by van Dokkum et al. (2006).

The rest-frame $\rfUV$ and $\rfVJ$ colors were determined by
interpolation between the directly observed bands using templates as a
guide.  For an in depth discussion of the algorithm, we refer the
reader to Rudnick et al. (2001; 2003).  We used an IDL implementation
of the algorithm by Taylor et al. (in preparation) dubbed
``InterRest''.

\subsubsection {Converting 24 $\mu$m flux to total IR luminosity}
\label{Lir.sec}
At redshifts $1.5<z<2.5$, the 24 $\mu$m fluxes trace the rest-frame
7.7 $\mu$m emission from polycyclic aromatic hydrocarbons (PAHs).  To
convert this MIR emission to a total IR luminosity $L_{\rm IR} \equiv
L(8-1000\ \mu$m$)$, we use the infrared spectral energy distributions of
star-forming galaxies provided by Dale \& Helou (2002).  The template
set allows us to quantify the IR/MIR flux ratio for different heating
levels of the interstellar environment, parameterized by $dM(U) \sim
U^{-\alpha} dU$ where M(U) represents the dust mass heated by an
intensity U of the interstellar radiation field.

We computed the total infrared luminosity $L_{\rm IR,\alpha}$ for each
object for all Dale \& Helou (2002) templates within the reasonable
range from $\alpha=1$ for active galaxies to $\alpha=2.5$ for
quiescent galaxies.  The mean of the resulting $\log (L_{\rm IR,\alpha=1,
1.0625, ..., 2.5})$ was adopted as best estimate for the IR
luminosity, and the $\pm 0.45$ dex variation from $L_{\rm IR,\alpha=2.5}$
to $L_{\rm IR,\alpha=1}$ was taken as a measure for the systematic
uncertainty in the conversion.

Apart from the random photometric error and systematic template
uncertainty, uncertainties in the photometric redshift contribute to
the total error budget.  For each galaxy, we calculated the spread in
$L_{\rm IR}$ caused by variations of $z_{\rm phot}$ within the 68\% confidence
interval.  Although the uncertainty in photometric redshift is partly
random (propagating from photometric uncertainties in the SED), we
treat it as purely systematic, originating from template mismatches.
This means the error bars related to $z_{\rm phot}$ on the stacked
$L_{\rm IR}$ measurements do not scale with $1/ \sqrt{N}$.  Instead, they
range from the stacked $L_{\rm IR}$ based on the lowest
$L_{\rm IR,individual}$ estimates allowed for each object within its
$z_{\rm phot}$ uncertainty, to the stacked $L_{\rm IR}$ based on the maximum
$L_{\rm IR,individual}$ allowed for each object.

\subsubsection {$L_{\rm IR}$ versus rest-frame color}
\label{Lir_rfcol.sec}

\begin {figure} [htbp]
\centering
\plotone{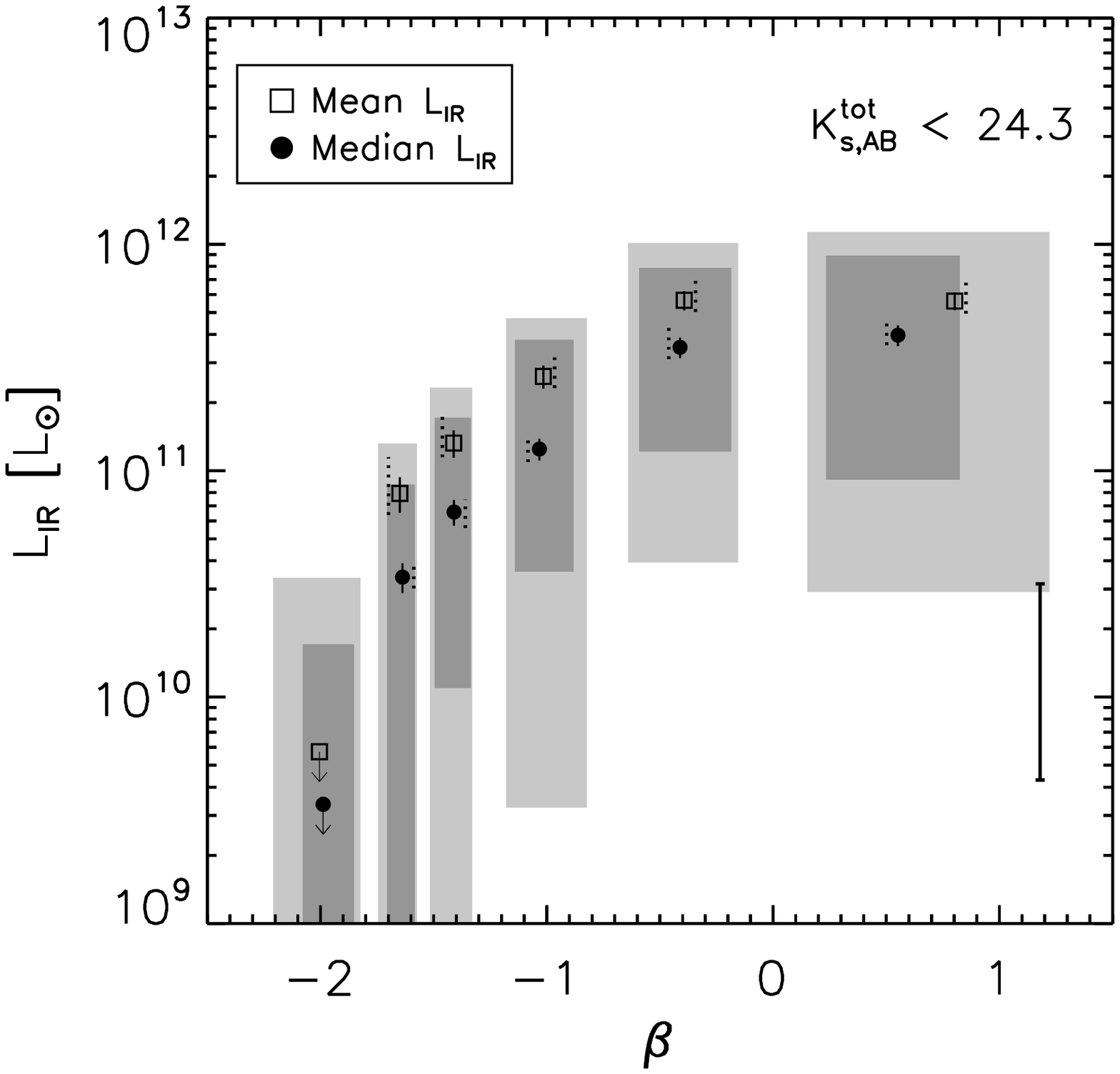}
\plotone{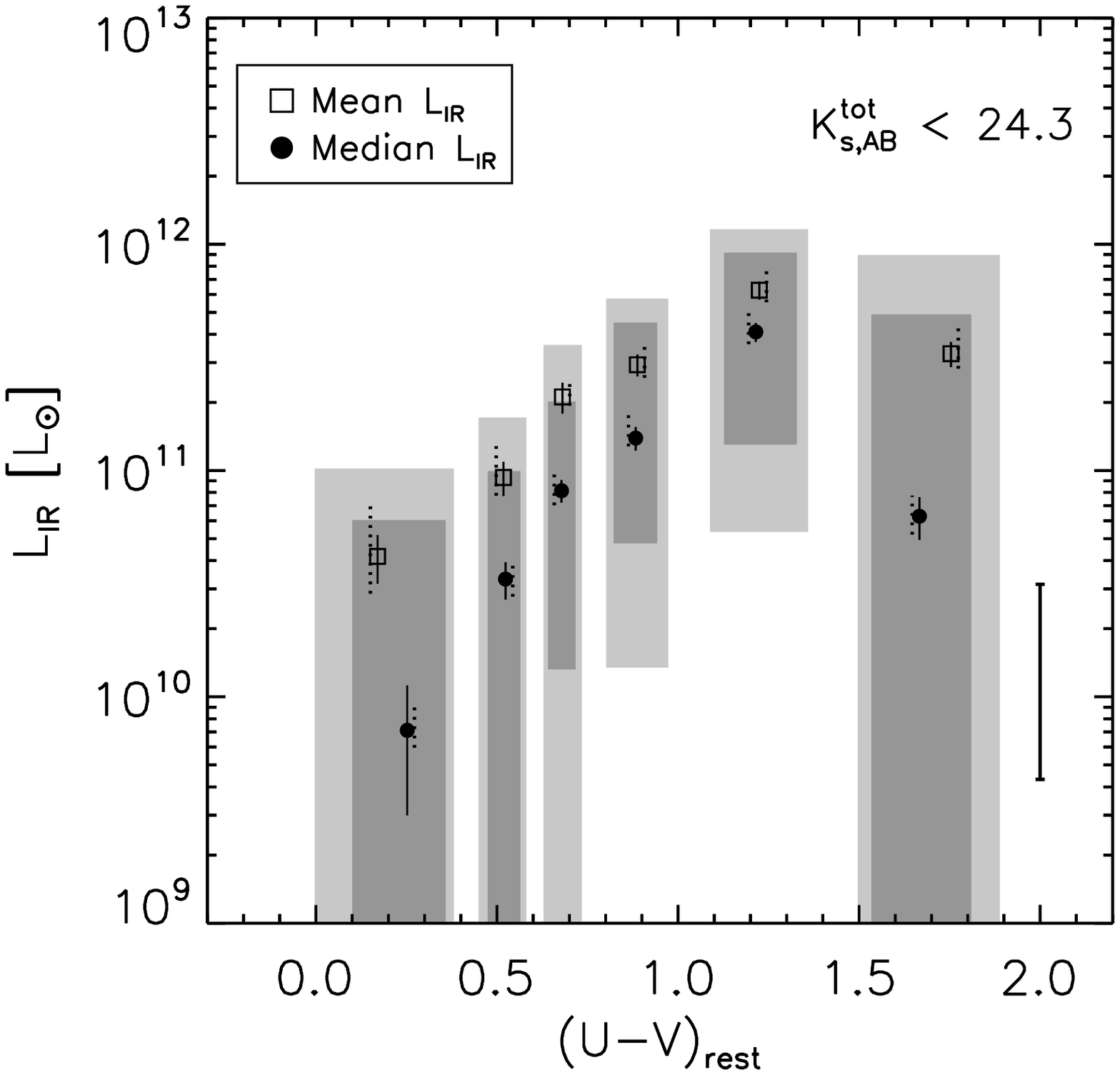}
\plotone{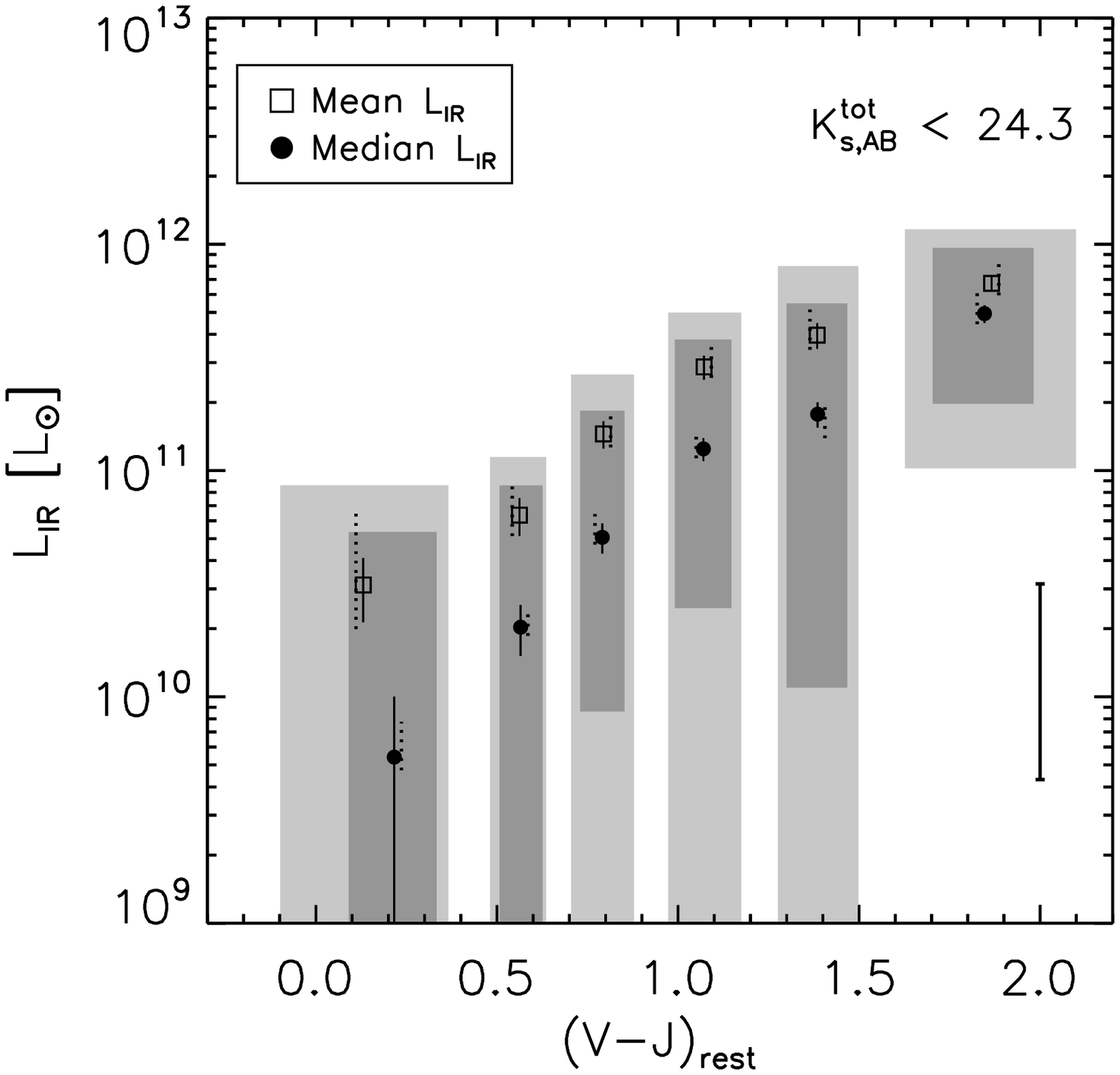}
\caption{\footnotesize Stacked total IR luminosities as function of
rest-frame UV slope, $\rfUV$, and $\rfVJ$ for non-AGN at $1.5<z<2.5$
with $S/N > 5$ in the $K_s$-band ($K_{s,{\rm AB}}^{\rm tot} \lesssim 24.3$).
Filled circles represent the median for each equal-number bin.  Empty
boxes represent the mean flux.  Light-grey and dark-grey polygons
indicate respectively the central 68\% and 50\% of the distribution
within each bin.  The systematic uncertainty induced by template
uncertainties in the conversion to $L_{\rm IR}$ is indicated in the bottom
right corner.  The dotted error bars indicate the variation in the
mean and median $L_{\rm IR}$ by systematic variations in $z_{\rm phot}$.  We
find a trend of increasing $L_{\rm IR}$ with redder rest-frame color.
Since the bins contain equal numbers of objects, this also means that
red galaxies contribute most to the integrated IR emission of distant
$K_s$-selected galaxies.
\label {Lir.fig}  
}
\end {figure}

Having determined the rest-frame colors and total IR luminosities of
$K_s$-selected galaxies without X-ray detection at $1.5<z<2.5$, we now
proceed to investigate which $K_s$-selected galaxies contribute most
to the IR emission.  Again, we investigate the average properties of
galaxies binned by color to enhance the robustness of our results.  In
Fig.\ \ref{Lir.fig}, we plot the mean and median total IR luminosities
of our in color bins divided sample versus the rest-frame UV slope
$\beta$, the rest-frame optical color $\rfUV$, and the rest-frame
optical-to-NIR color $\rfVJ$.  The black error bars indicate the error
in the mean, respectively median.  With dotted error bars, we show the
systematic variations allowed within the photometric redshift
uncertainties.  Finally, the error bar in the bottom right corner
represents the range from quiescent to active galaxy templates by Dale
\& Helou (2002).  Clearly, systematic uncertainties are dominating the
error budget in this analysis.

As in Fig.\ \ref{MIPSobs.fig}, we find a large range in IR properties
within each bin, illustrated by the light grey and dark grey polygons
that mark respectively the central 68\% and 50\% of the distribution
of $L_{\rm IR}$ of individual objects.  Nevertheless, a general trend is
visible of redder colors corresponding to larger IR luminosities.  No
matter which part of the spectral energy distribution is used to
define red or blue galaxies, the redder half of the galaxies in our
sample have stacked IR properties in the LIRG ($L_{\rm IR} = 10^{11} -
10^{12}\ L_{\sun}$) regime.  The trend seems to flatten at the reddest
UV slopes, and the contribution to the total IR luminosity even drops
for the reddest $\rfUV$ bin.  In $\rfVJ$ on the other hand, the
increase in $L_{\rm IR}$ with reddening color continues over the entire
color range, reaching ULIRG luminosities ($L_{\rm IR} > 10^{12}\
L_{\sun}$) for 23\% of the galaxies in the reddest bin.

Considering the IR luminosities of individual objects over the entire
color range, we find that ULIRGs make up 7\% of our $K_s$-selected
sample with $K^{\rm tot}_{s,{\rm AB}} < 24.3$.  The fraction of ULIRGs increases
to 35\% when only considering massive ($M > 10^{11}\ M_{\sun}$)
galaxies.  A similar fraction was found by Daddi et
al. (2007).

The IR/MIR conversion factor varies by nearly an order of magnitude
between the use of quiescent ($\alpha=2.5$) or active ($\alpha=1$)
galaxy templates.  For each of the galaxies, we used the mean of the
logarithm of all conversion factors derived from the $1 \leq \alpha \leq
2.5$ templates.  Other authors (e.g. Papovich et al. 2006; Papovich et
al. 2007) adopted a conversion to total IR luminosity in which the
rest-frame IR luminosity density translates uniquely to a single SED
template (i.e., a lower $\alpha$ was used for brighter galaxies).
Applying the conversion by Papovich et al. (2007) to each of the
galaxies in our sample and repeating the stacking procedure, gives the
following results.  The mean $L_{\rm IR}$ for each color bin increases by
typically $\sim 0.1$ dex with respect to the value obtained with our
universal (flux-independent) conversion.  The median $L_{\rm IR}$ for each
color bin instead decreases by typically $\sim 0.15$ dex.  All the
described correlations between $L_{\rm IR}$ and rest-frame color remain
intact.  We thus confirm that our results are robust to alternative
conversions from $24\ \mu$m flux to total IR luminosity that are
commonly used in the literature.

We conclude that amongst distant $K_s$-selected galaxies that show no
sign of relatively unobscured AGN at X-ray wavelengths, the redder
galaxies have on average larger total IR luminosities.  Given that
each bin in Fig.\
\ref{Lir.fig} contains an equal number of objects, it is also clear
that red galaxies in our sample contribute more to the integrated
total IR luminosity than blue galaxies.  We argue that this trend
cannot be explained by systematic errors, and tested that this
conclusion is robust against the precise choice of redshift interval
by varying the lower edge between redshift 1 and 2 and the upper edge
between 2 and 3.  Likewise, we verified that none of our conclusions
critically depend on the number of color bins.  Finally, we note that,
although X-ray detections were excluded from our analysis to validate
the use of IR templates for starforming galaxies, the results remain
nearly unaffected when we treat them as normal galaxies.

\begin {figure} [htbp]
\centering
\plotone{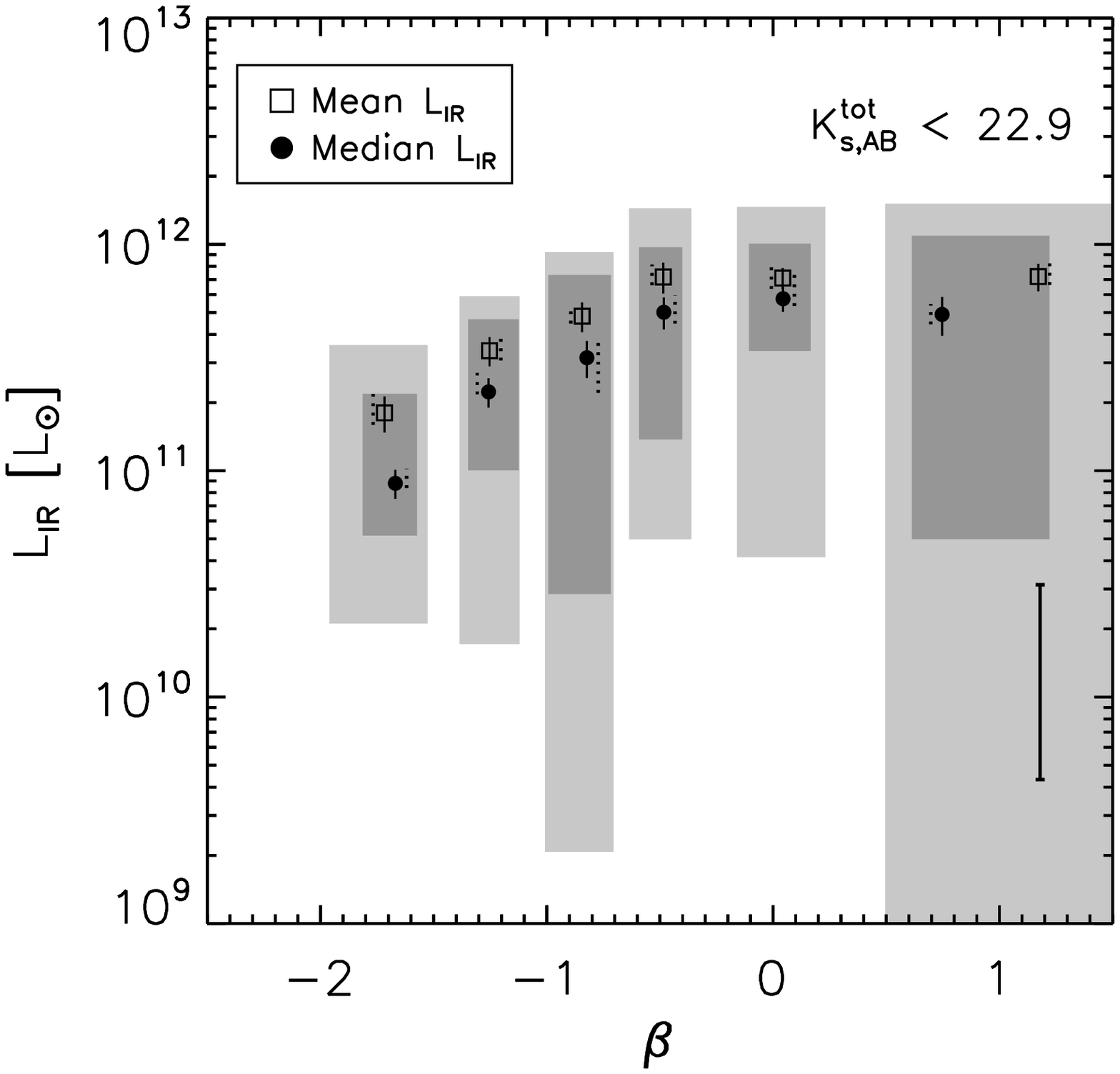}
\plotone{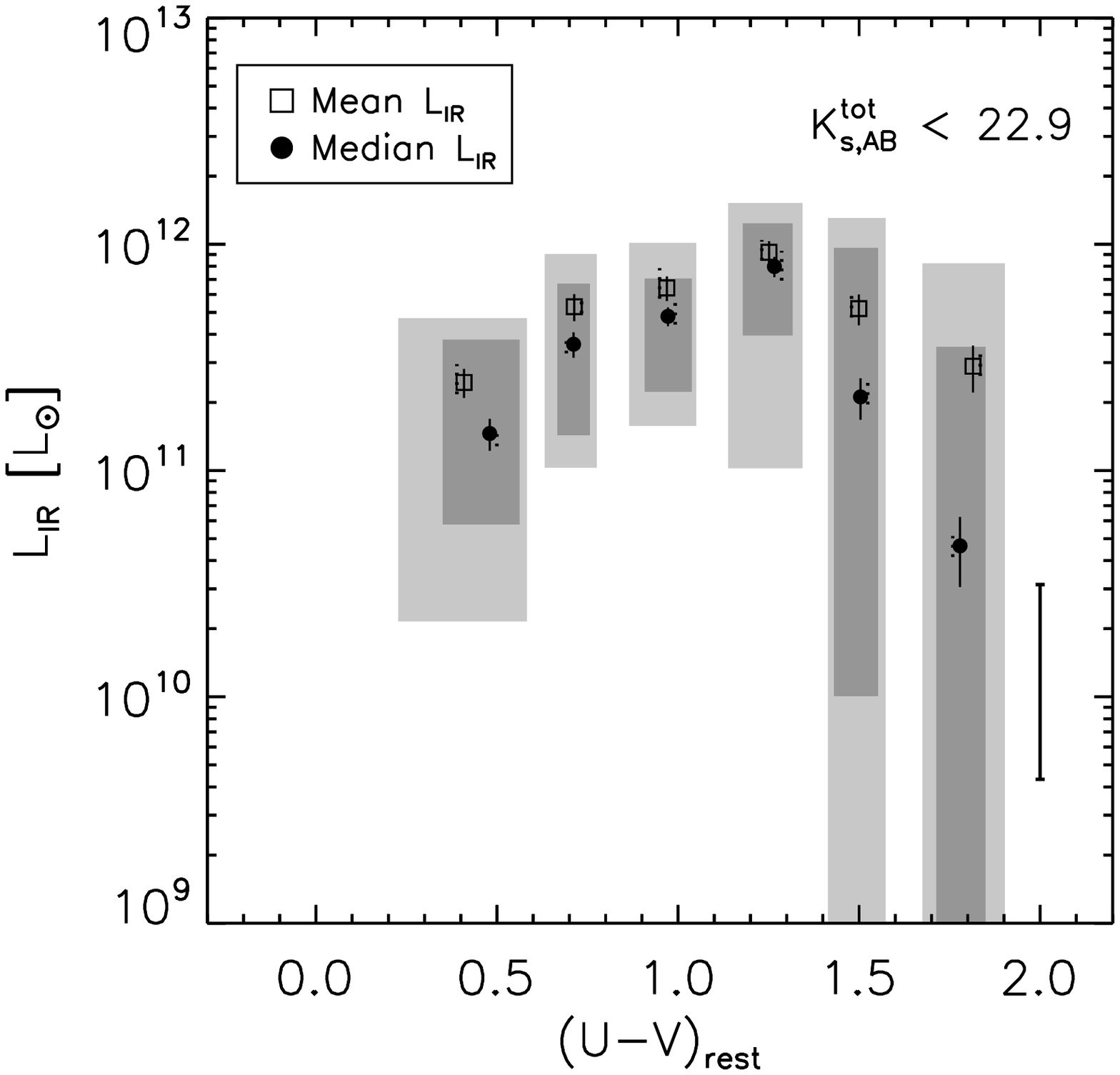}
\plotone{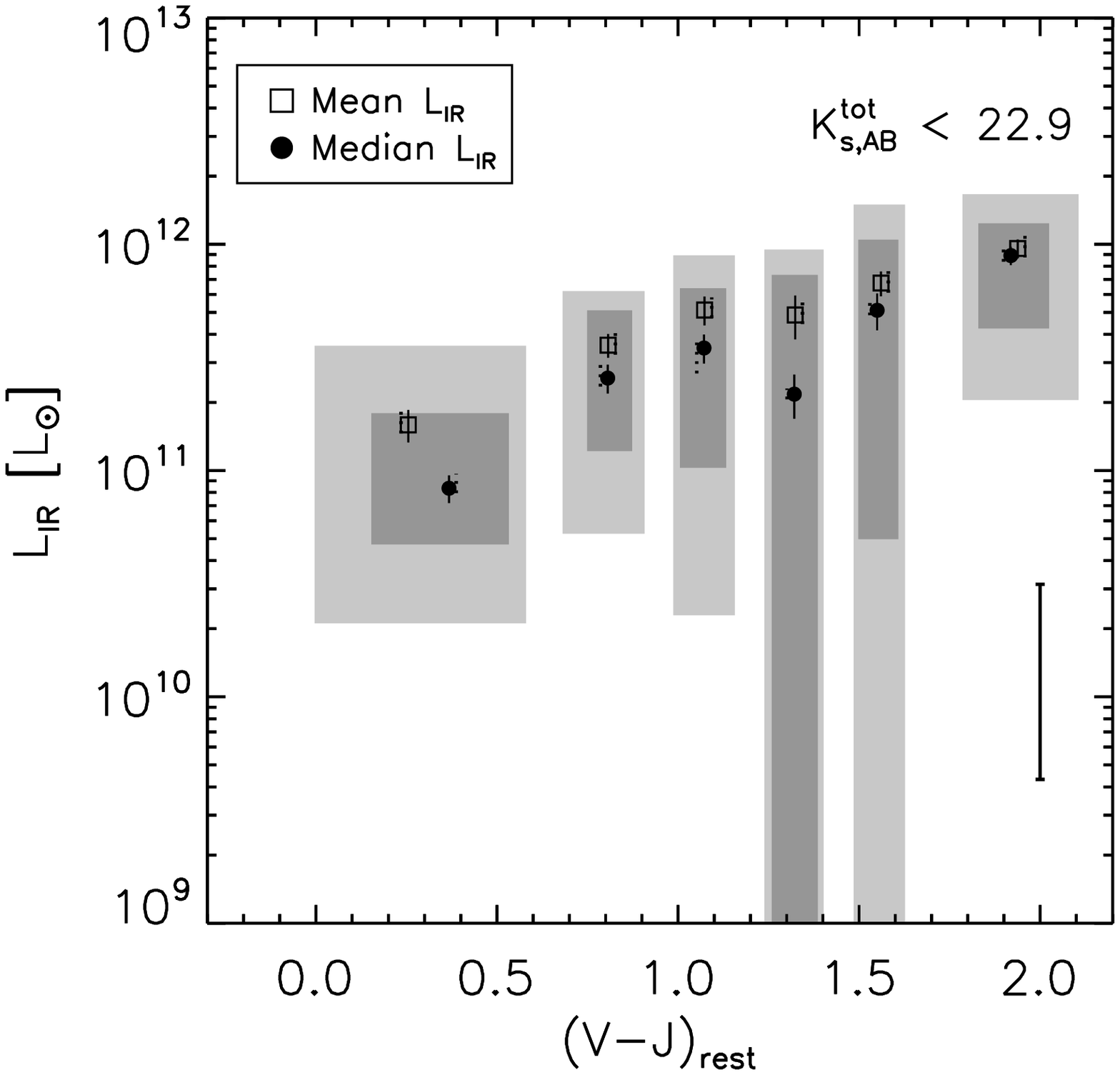}
\caption{\footnotesize Stacked total IR luminosities as function of
rest-frame UV slope, $\rfUV$, and $\rfVJ$ for non-AGN at $1.5<z<2.5$
with $K_{s,{\rm AB}}^{\rm tot} \lesssim 22.86$ ($K_{s,{\rm Vega}}^{\rm
tot} \lesssim 21$).  Filled circles represent the median for each
equal-number bin.  Empty boxes represent the mean flux.  Light-grey
and dark-grey polygons indicate respectively the central 68\% and 50\%
of the distribution within each bin.  The systematic uncertainty
induced by template uncertainties in the conversion to $L_{\rm IR}$ is
indicated in the bottom right corner.  The dotted error bars indicate
the variation in the mean and median $L_{\rm IR}$ by systematic variations
in $z_{\rm phot}$.  The increase in $L_{\rm IR}$ with redder rest-frame color
is less pronounced than for the fainter ($K_{s,{\rm AB}}^{\rm tot} \lesssim
24.3$, Fig.\
\ref{Lir.fig}) sample.  The drop in $L_{\rm IR}$ at the reddest $\rfUV$
colors is stronger.
\label {Lir_Kbright.fig}  
}
\end {figure}

Dividing the sample in two equal-number bins according to $\beta$,
$\rfUV$, and $\rfVJ$, we find that the integrated total IR luminosity
of the red half is larger than that of the blue half by a factor $6.7
\pm 1.6$, $3.6 \pm 0.8$, and $5.6 \pm 1.2$ respectively.  Imposing a
brighter cut in the $K_s$-band magnitude (Fig.\
\ref{Lir_Kbright.fig}) weakens the correlation of $L_{\rm IR}$ with UV
slope and $\rfVJ$.  Adopting a $K^{\rm tot}_{s,{\rm AB}} < 22.86$ cut, as is the
case for the NIR-selected sample studied by Reddy et al. (2006),causes
the stacked $L_{\rm IR}$ of the bluest $\rfUV$ colors to increase while
the stacked $L_{\rm IR}$ of the reddest $\rfUV$ colors drops.  The overall
fraction of galaxies that is not detected at the 3$\sigma$ level in
the 24 $\mu$m band is lower by a factor 2 in the $K^{\rm tot}_{s,{\rm AB}} <
22.86$ sample than in our $K^{\rm tot}_{s,{\rm AB}} < 24.3$ sample.  However, at
the reddest optical colors ($\rfUV > 1.3$) the fraction of galaxies
that is not detected at the 3$\sigma$ level in the 24 $\mu$m band
remains the same as we impose the brighter $K_s$-band limit.  We
find that, for this $K_s$-bright sample, the ratio of the integrated
$L_{\rm IR}$ of the red and the blue half of the galaxies in $\beta$,
$\rfUV$, and $\rfVJ$ amounts to a factor of $2.1 \pm 0.5$, $1.2 \pm
1.1$, and $2.1 \pm 0.5$ respectively.

It is tempting to elaborate on the physical interpretation in terms of
star formation rate (SFR), age, and dust content of the galaxies in
our sample implied by the presented results.  Colors in different
wavelength regimes are to a greater or lesser extent determined by
these physical parameters.  The UV slope and $\rfVJ$ color are both
sensitive tracers of dust attenuation (Meurer et al. 1999 and Wuyts et
al. 2007 respectively).  In comparison to the $\rfVJ$ color, the
$\rfUV$ color is more sensitive to stellar age and to lesser extent
affected by dust (Wuyts et al. 2007).  The fact that in the reddest
$\rfUV$ bin the total IR luminosity drops again might therefore
indicate an increasing contribution from low $L_{\rm IR}$ galaxies with
little dust-obscured star formation.  In combination with the fact
that rest-frame optically selected galaxies often have faint UV
luminosities and thus little unobscured star formation (F\"orster
Schreiber et al. 2004), this suggests that part of the galaxies making
up the reddest $\rfUV$ bin have a low overall SFR (unobscured +
obscured).  This is consistent with the spectral evidence for galaxies
with quenched star formation at $z \sim 2$ (Kriek et al. 2006).  A
similar conclusion was drawn by Reddy et al. (2006), who found for a
sample of galaxies at similar redshifts selected by optical and NIR
color criteria that the IR luminosity of 24 $\mu$m detected sources
increased toward redder observed $z-K$, but that at the reddest $z-K$
colors a population without 24 $\mu$m detection exists that satisfies
the Distant Red Galaxy (Franx et al. 2003) and/or BzK/PE (Daddi et
al. 2004) color criteria.  Based on X-ray stacking in the
GOODS-North field, although also probing only to $K_{s,{\rm AB}}^{\rm tot} <
22.86$, Reddy et al. (2005) found a similar turnover in inferred SFR
at $z-K>3$.  Although the observed $z-K$ color is redshift dependent
and at $z \sim 2$ spans a somewhat broader wavelength range than
$\rfUV$, both colors probe the age-sensitive Balmer/4000\AA\ break and
the observed turnovers therefore likely share the same origin.  

An in depth analysis of the mix between dust-obscured starforming
systems and evolved red galaxies would require a careful SED modeling,
estimating the SFR based on different wavelength tracers from X-ray
over UV to the infrared, and a treatment of each object on an
individual basis to assess their relative contribution.  Such a study
is clearly beyond the scope of this paper, and will be presented by
Labb\'{e} et al. (in preparation) based on the combined sample of
galaxies in the CDFS (this paper),
\1054 (FS06) and the HDFS (L03).

\subsubsection {Redshift dependence}
\label{depend_z.sec}

We measure a median rest-frame color evolution for galaxies in our
$K_{s,{\rm AB}}^{\rm tot} < 24.3$ sample from $z=2.5$ to $z=1.5$ of
0.48, -0.05, and 0.48 for $\beta$, $\rfUV$, and $\rfVJ$ respectively.
We note that going to lower redshifts, the evolution in $\beta$ and
$\rfVJ$ reverses.  The observed evolution amounts to 26, 5, and 40\%
of the scatter in the respective rest-frame color at any given
redshift in the $1.5<z<2.5$ range.  Consequently, not every color bin
in Fig.\ \ref{Lir.fig} and Fig.\ \ref{Lir_Kbright.fig} is composed of
galaxies with exactly the same redshift distribution.  This is
particularly the case for $\beta$ and $\rfVJ$, where the measured
color evolution with redshift is largest.  A dependence on redshift
can potentially affect the observed trend of increasing $L_{\rm IR}$
toward redder colors, if $L_{\rm IR}$ evolves within the $1.5<z<2.5$
interval.  Binning galaxies by redshift, we find a decrease in the
stacked $L_{\rm IR}$ from $z=2.5$ to $z=1.5$ by $\sim 0.4$ dex.  We
conclude that redshift dependencies within the $1.5<z<2.5$ interval
are not driving, and if anything weaken, the observed increase of
$L_{\rm IR}$ toward redder colors.

\subsubsection {Mass dependence}
\label{depend_M.sec}

We now investigate the role of stellar mass in the observed trend of
red galaxies dominating the total IR emission.  van Dokkum et
al. (2006) showed that the high-mass end of the high redshift galaxy
population is dominated by galaxies with red $\rfUV$ and large
$\beta$.  Considering the optical-to-NIR colors of galaxies in our
sample, a similar lack of massive galaxies with blue $\rfVJ$ is
observed.

\begin {figure*} [htbp]
\centering
\plottwo{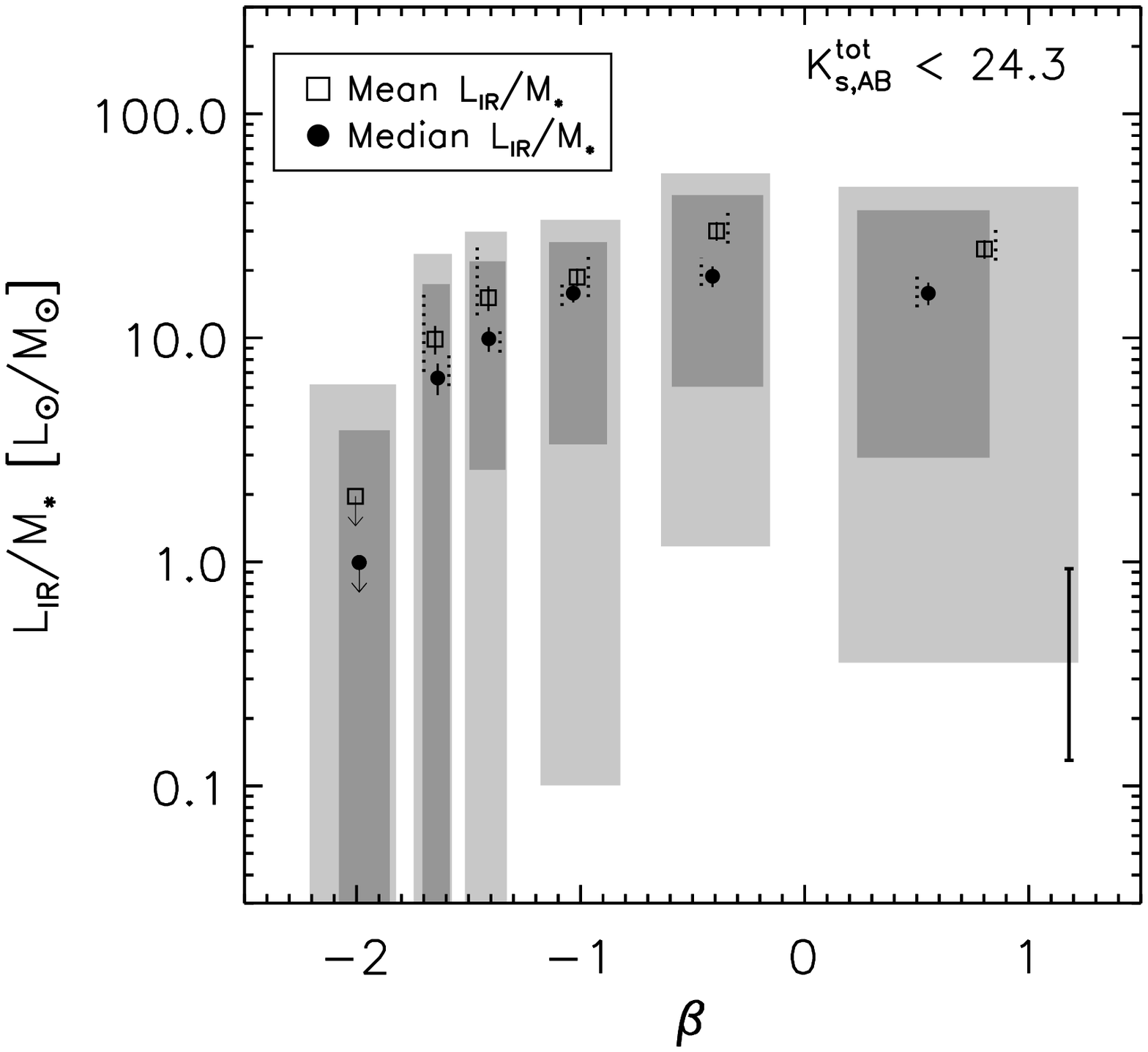}{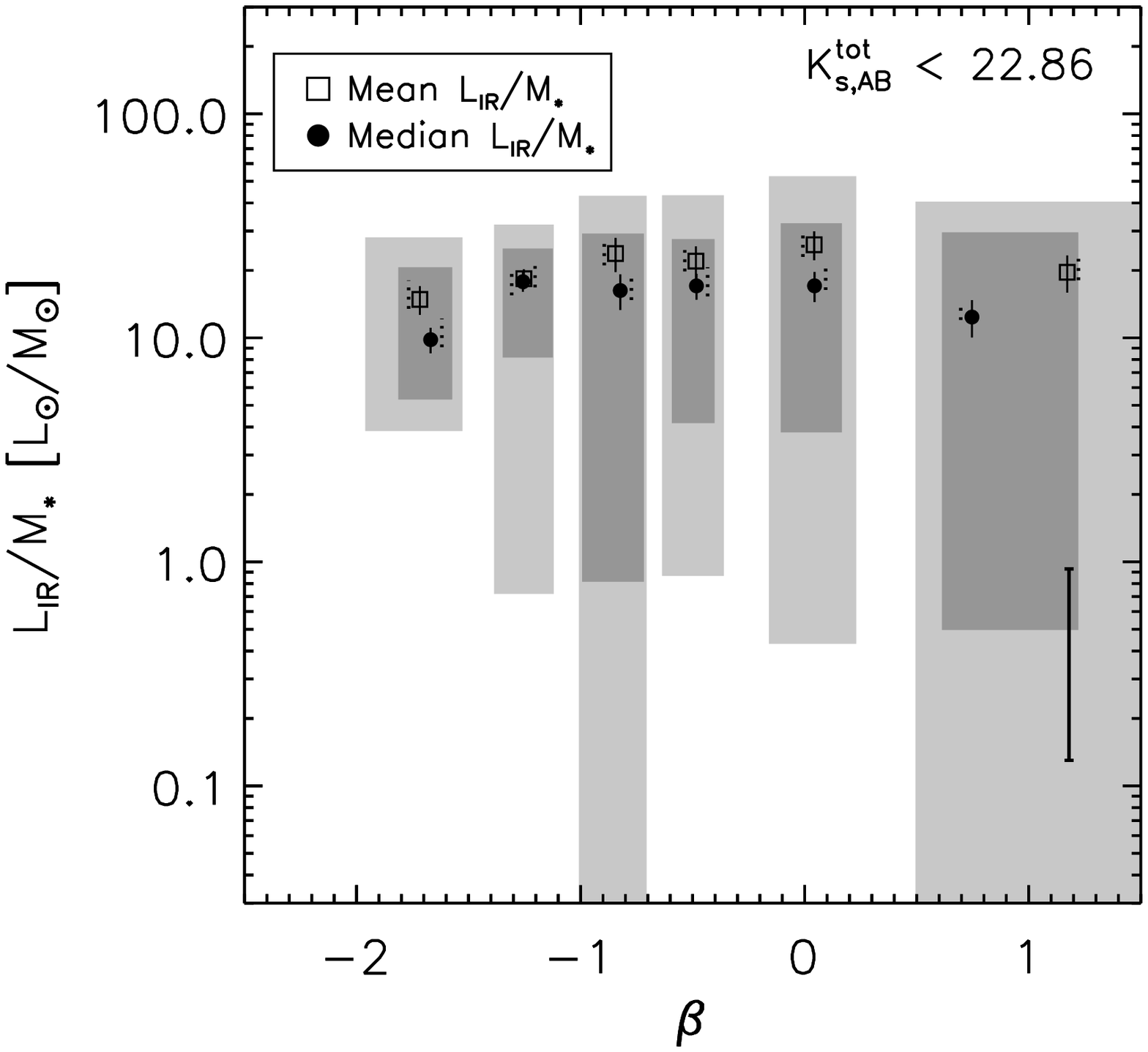}
\plottwo{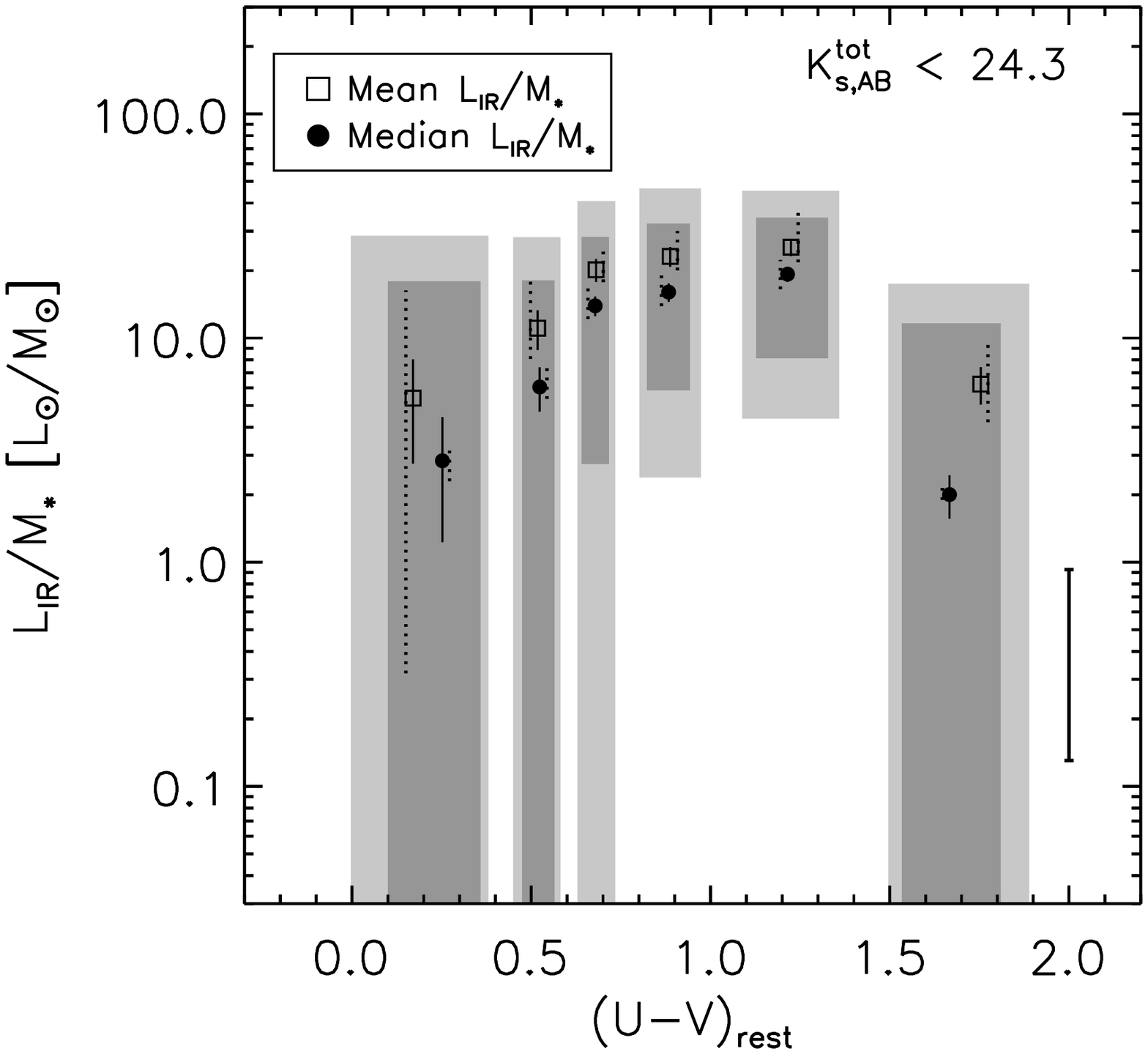}{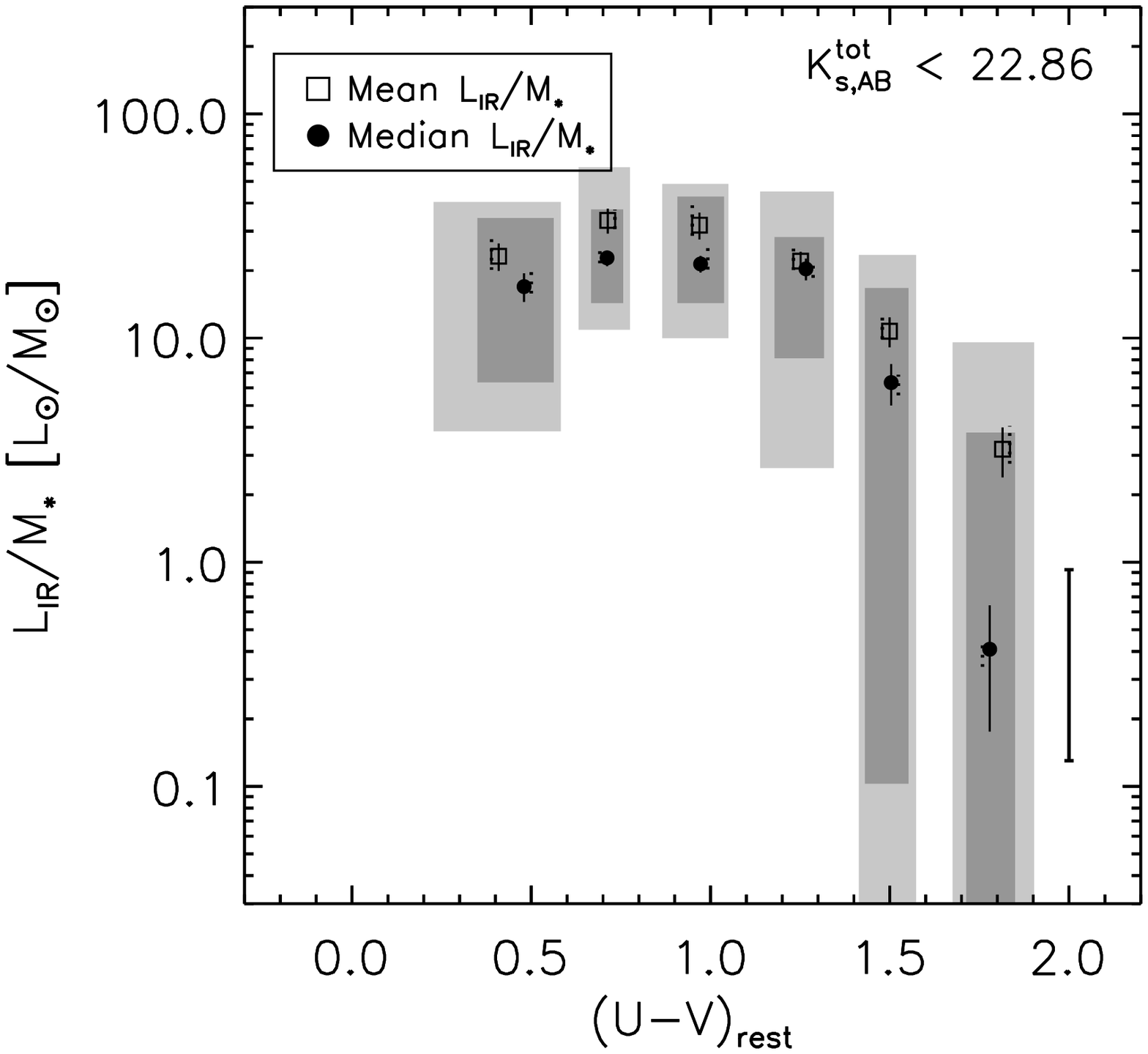}
\plottwo{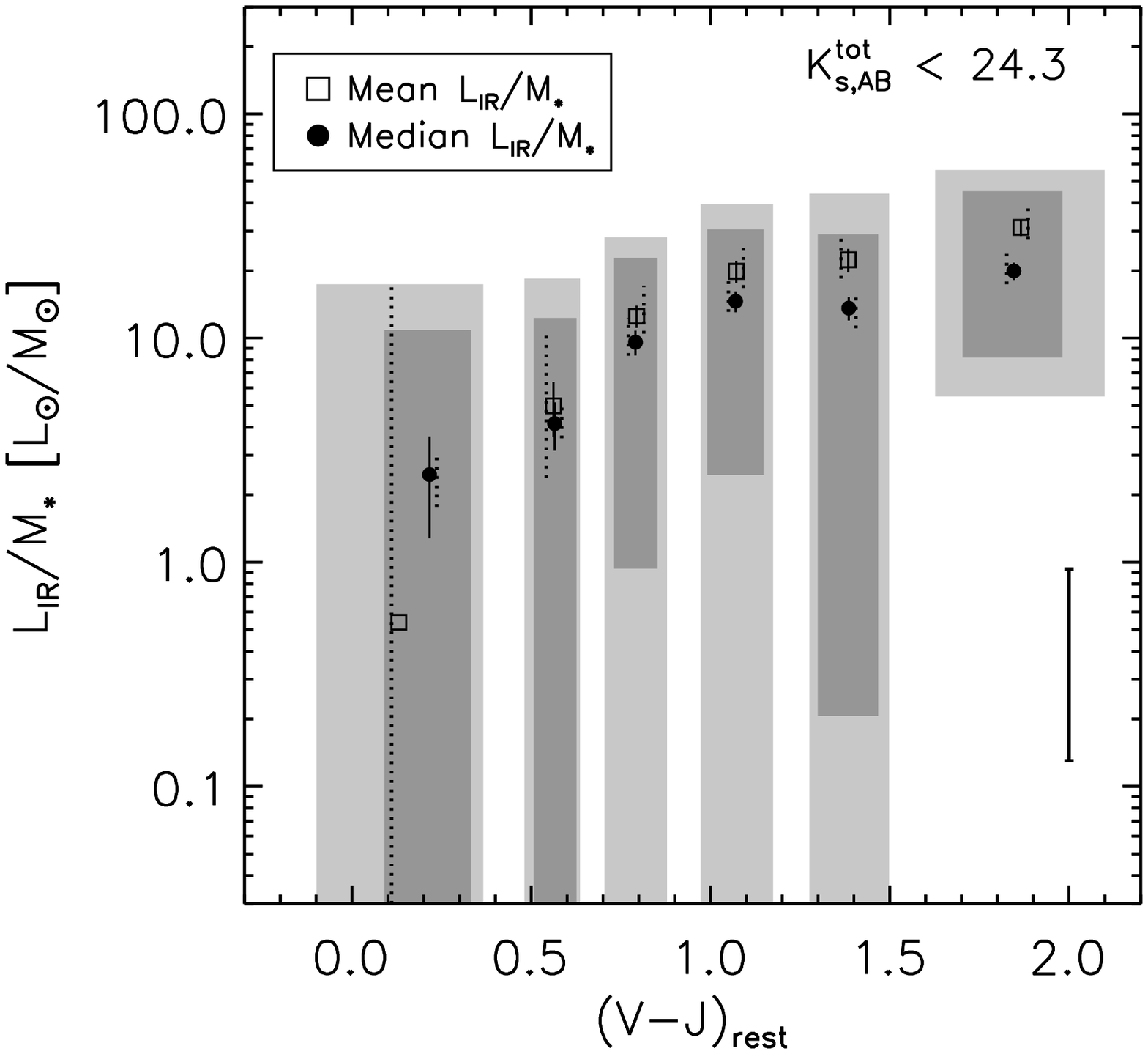}{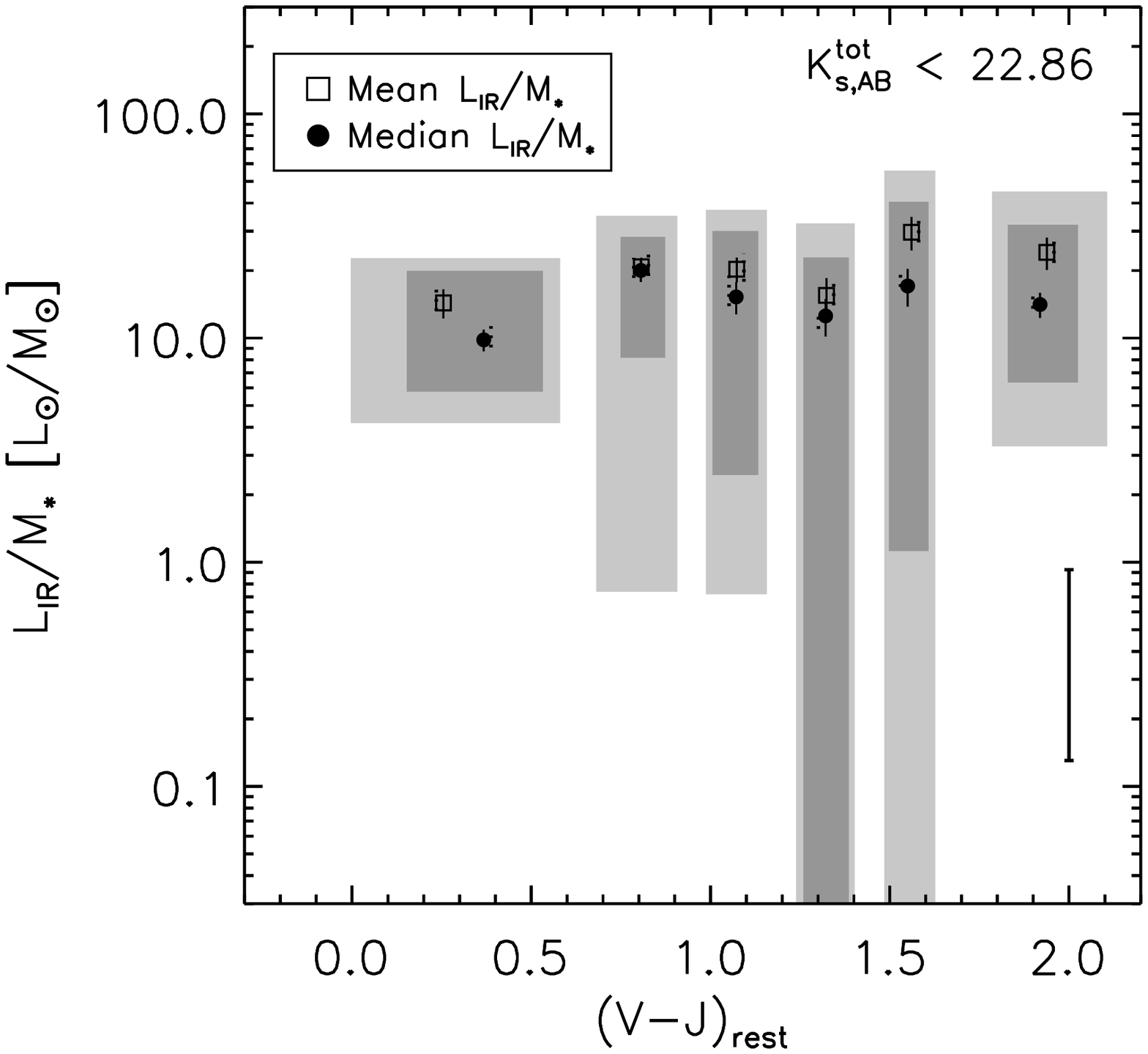}
\caption{\footnotesize Stacked total IR luminosity divided by stellar mass as function of
rest-frame UV slope, $\rfUV$, and $\rfVJ$ for non-AGN at $1.5<z<2.5$
with $K_{s,{\rm AB}}^{\rm tot} \lesssim 24.3$ ({\it left panels}) and
$K_{s,{\rm AB}}^{\rm tot} \lesssim 22.86$ ({\it right panels}).
Symbols are identical to those in Fig.\ \ref{Lir.fig} and Fig.\
\ref{Lir_Kbright.fig}.  The stacked $L_{\rm IR}/M_*$ shows a weaker
increase with rest-frame color than the stacked $L_{\rm IR}$ (Fig.\
\ref{Lir.fig} and Fig.\ \ref{Lir_Kbright.fig}), and is nearly
independent of $\beta$ and $\rfVJ$ for the $K_{s,{\rm AB}}^{\rm tot} <
22.86$ subsample.  A drop in the stacked $L_{\rm IR}/M_*$ at the
reddest $\rfUV$ is clearly observed.
\label {Lir_over_M.fig}  
}
\end {figure*}

In order to investigate whether the $L_{\rm IR}$ - color relation
discussed in \S\ref{Lir_rfcol.sec} merely reflects a correlation
between stellar mass and $L_{\rm IR}$, we now divide out the mass
dependence.  Fig.\ \ref{Lir_over_M.fig} shows the stacked $L_{\rm IR}
/ M_*$ measurements as a function of color for the $K_{s,{\rm
AB}}^{\rm tot} < 24.3$ sample and the $K_{s,{\rm AB}}^{\rm tot} <
22.86$ subsample.  The y-axes span 4 orders of magnitude, as in Fig.\
\ref{Lir.fig} and Fig.\ \ref{Lir_Kbright.fig}, allowing an easy visual
comparison of the strengths of the trends.  A weaker trend, by up to
an order of magnitude, is observed between $L_{\rm IR} / M_*$ and
rest-frame color than between $L_{\rm IR}$ and rest-frame color (Fig.\
\ref{Lir.fig} and Fig.\ \ref{Lir_Kbright.fig}).  For the $K_s$-bright subsample ($K_{s,{\rm AB}}^{\rm tot} <
22.86$), the observed $L_{\rm IR}$ versus $\beta$ and $L_{\rm IR}$
versus $\rfVJ$ relation can even be entirely attributed to a
dependence of color on stellar mass.  Finally, the drop in stacked
$L_{\rm IR} / M_*$ at the reddest $\rfUV$ colors is stronger than the
drop in stacked $L_{\rm IR}$.  The population of optically red
galaxies with only little IR emission, suggested by the analysis in
\S\ref{Lir_rfcol.sec}, therefore concerns galaxies at the high-mass
end of the mass function.

\section {Summary}
\label{summary.sec}
We present a $K_s$-band selected catalog for the GOODS-CDFS, dubbed
FIREWORKS, containing consistent photometry in the $U_{38}$,
$B_{435}$, $B$, $V$, $V_{606}$, $R$, $i_{775}$, $I$, $z_{850}$, $J$,
$H$, $K_s$, $[3.6\mu$m$]$, $[4.5\mu$m$]$, $[5.8\mu$m$]$,
$[8.0\mu$m$]$, and $[24\mu$m$]$ bands.  Together with the photometry,
we release a list of photometric redshifts with a scatter in $\Delta z
/ (1+z)$ of 0.032, a cross-correlation with all available
spectroscopic redshifts to date, and a cross-correlation with the 1Ms
X-ray catalog by Giacconi et al. (2002).  After a description of the
catalog construction, we discuss the differences with the GOODS-MUSIC
'$z_{850}$ + $K_s$'-selected catalog by G06a.

A previous version of the catalog, lacking the WFI $U_{38}BVRI$
photometry, has been used to estimate stellar mass densities (Rudnick
et al. 2006), construct luminosity functions (Marchesini et al. 2006)
and study the predominance of red galaxies at the high mass end (van
Dokkum et al. 2006).  In this paper, we exploit the full catalog to
answer the following question: Which distant $K_s$-band selected
galaxies are brightest and contribute most to the total IR luminosity?

First, we compared the stacked 24 $\mu$m fluxes of galaxies at
$1.5<z<2.5$ with $K_{s,{\rm AB}}^{\rm tot} < 24.3$ split in observed color bins.
Overall, a large spread in IR properties is found in each color bin.
Nevertheless, stacking the fluxes within each bin reveals a clear
trend with color.  Both in the observed $B_{435}-V_{606}$, $J-K_s$, and $K_s -
[4.5\mu$m$]$ colors, the lowest mean and median $[24\mu$m$]_{\rm tot}$
fluxes are found for the bluest color bin.  In $J-K_s$ and $K_s -
[4.5\mu$m$]$, the emission at 24 $\mu$m continues to rise toward redder
colors.

Second, we use our photometric redshifts to convert the observed
spectral energy distributions to rest-frame colors and translate the
observed 24 $\mu$m flux to the total IR luminosity $L_{\rm IR} \equiv
L(8-1000\ \mu$m$)$.  In this procedure, all relatively unobscured AGN
candidates, selected by their X-ray detection, were rejected from the
sample.  Removing the redshift dependence and extrapolating from MIR
to total IR comes at the cost of systematic uncertainties.  We
carefully measured the systematic contribution to the total error
budget from uncertainties in $z_{\rm phot}$ and from our lack of knowledge
about which IR template SED matches best the spectral shape of the
objects in our sample.  Doing so, we find a continuous increase in
$L_{\rm IR}$ with $\rfVJ$.  An increasing $L_{\rm IR}$ is also measured with
UV slope $\beta$, flattening at the largest $\beta$.  The rising trend
of the stacked $L_{\rm IR}$ luminosity toward redder $\rfUV$ seems to
reverse in the reddest color bin.  The large range of total IR
properties in this bin suggests a mixture of galaxies with large
amounts of dust emission (LIRGs up to ULIRGs) and objects devoid of
it.  We note that, if we were to apply a different translation from
MIR to total IR luminosity than simply averaging over the conversion
factors derived from all reasonable templates, the observed trends
remain intact.  This is e.g. the case when an SED template
corresponding to a larger heating intensity of the interstellar
radiation field is used for objects with a larger rest-frame IR
luminosity density, as done by Papovich et al. (2006; 2007).

Since we divide our $K_s$-band selected sample in bins containing
equal numbers of objects, it is immediately clear that not only do red
galaxies have on average the largest total IR luminosities, it is also
true that they form the dominant contribution to the overall total IR
luminosity emitted by $K_s$-selected galaxies at $1.5<z<2.5$.  We show
that the dependence of both color and $L_{\rm IR}$ on stellar mass is
an important driver of, but cannot fully explain the observed trends
of $L_{\rm IR}$ with color.

\acknowledgements
We thank Adriano Fontana and Andrea Grazian for useful discussions
about the photometric aspect of this work.  SW acknowledges support
from the W. M. Keck Foundation.  Support from NSF CAREER grant AST
04-49678 is gratefully acknowledged.  Observations have been carried
out using the Very Large Telescope at the ESO Paranal Observatory
under Program IDs LP168.A-0485, 170.A-0788,074.A-0709, 275.A-5060, and
171.A-3045.

\begin{references}
{\small
\reference{} Arnouts, S., et al. 2001, A\&A, 379, 740
\reference{} Beckwith, S. V. W., et al. 2006, AJ, 132, 1729
\reference{} Bertin, E.,\& Arnouts, S. 1996, A\&AS, 117, 393
\reference{} Bruzual, G.,\& Charlot, S. 2003, MNRAS, 344, 1000 (BC03)
\reference{} Calzetti, D., Kinney, A. L.,\& Storchi-Bergmann, T. 1994, ApJ, 429, 582
\reference{} Calzetti, D., et al. 2000, ApJ, 533, 682
\reference{} Caputi, K. I., Dole, H., Lagache, G., McLure, R. J., Dunlop, J. S., Puget, J.-L., Le Floc'h, E.,\& P\'{e}rez-Gonzal\'{e}z, P. G. 2006, A\&A, 454, 143
\reference{} Chary, R.,\& Elbaz, D. 2001, ApJ, 556, 562        
\reference{} Cimatti, A., et al. 2002, A\&A, 381, L68
\reference{} Cristiani, S., et al. 2000, A\&A, 359, 489
\reference{} Croom, S. M., Warren, S. J.,\& Glazebrook, K. 2001, MNRAS, 328, 150
\reference{} Daddi, E., Cimatti, A., Renzini, A., Fontana, A., Mignoli, M., Pozzetti, L., Tozzi, P.,\& Zamorani, G. 2004, ApJ, 617, 746
\reference{} Daddi, E., et al. 2007, ApJ, 670, 156 
\reference{} Dale, D. A.,\& Helou, G. 2002, ApJ, 576, 159
\reference{} Doherty, M., Bunker, A. J., Ellis, R. S.,\& McCarthy, P. J. 2005, MNRAS, 361, 525
\reference{} Erben, T., et al. 2005, AN, 326, 432
\reference{} Fazio, G. G., et al. 2004, ApJS, 154, 10
\reference{} Fioc, M.,\& Rocca-Volmerange, B. 1997, A\&A, 326, 950
\reference{} F\"{o}rster Schreiber, N. M., et al. 2004, ApJ, 616, 40
\reference{} F\"{o}rster Schreiber, N. M., et al. 2006, AJ, 131, 1891
\reference{} Franx, M., et al. 2003, ApJ, 587, L79
\reference{} Giacconi, R., et al. 2000, A\&AS, 197, 9001
\reference{} Giacconi, R., et al. 2002, ApJS, 139, 369
\reference{} Giavalisco, M.,\& the GOODS Team 2004, ApJ, 600, L93
\reference{} Grazian, A., et al. 2006a, A\&A, 449, 951 (G06a)
\reference{} Grazian, A., et al. 2006b, A\&A, 453, 507
\reference{} Grazian, A., et al. 2007, A\&A, 465, 393
\reference{} Groenewegen, M. A. T., et al. 2002, A\&A, 392, 741
\reference{} Hildebrandt, H., et al. 2006, A\&A, 452, 1121
\reference{} Hopkins, A. M.,\& Beacom, J. F. 2006, ApJ, 651, 142
\reference{} Kriek, M., et al. 2006, ApJ, 645, 44
\reference{} Kriek, M., et al. 2007, ApJ, 669, 776 
\reference{} Kriek, M., et al. 2008, ApJ, in press (astro-ph/08011110)
\reference{} Labb\'{e}, I., et al. 2003, AJ, 125, 1107
\reference{} Le F\`evre, O., et al. 2004, A\&A, 428, 1043
\reference{} Marchesini, D., et al. 2007, 656, 42
\reference{} Meurer, G., Heckman, T. M.,\& Calzetti, D. 1999, ApJ, 521, 64
\reference{} Mignoli, M., Cimatti, A., Zamorani, G., et al. 2005, A\&A, 437, 883
\reference{} Norman, C., et al. 2002, ApJ, 571, 218
\reference{} Papovich, C., Dickinson, M., Giavalisco, M., Conselice, C. J.,\& Ferguson, H. C. 2005, ApJ, 631, 101
\reference{} Papovich, C., et al. 2006, ApJ, 640, 92
\reference{} Papovich, C., et al. 2007, 668, 45
\reference{} P\'{e}rez-Gonz\'{a}lez, P. G., et al. 2005, ApJ, 630, 82
\reference{} P\'{e}rez-Gonz\'{a}lez, P. G., et al. 2008, ApJ, 657, 234
\reference{} Persson, S. E., Murphy, D. C., Krzeminski, W., Roth, M.,\& Rieke, M. J. 1998, AJ, 116, 2475
\reference{} Popesso, P., et al. 2008, astro-ph/08022930
\reference{} Ravikumar, C. D., et al. 2007, A\&A, 465, 1099
\reference{} Reddy, N. A., et al. 2005, ApJ, 633, 748
\reference{} Reddy, N. A., et al. 2006, ApJ, 644, 792
\reference{} Rieke, G. H., et al. 2004, ApJS, 154, 25
\reference{} Roche, N. D., Dunlop, J.,\& Almaini, O. 2003, MNRAS, 346, 803
\reference{} Roche, N. D., Dunlop, J., Caputi, K. I., McLure, R., Willott, C. J.,\& Crampton, D. 2006, MNRAS, 370, 74
\reference{} Rudnick, G., et al. 2001, AJ, 122, 2205
\reference{} Rudnick, G., et al. 2003, ApJ, 599, 847
\reference{} Rudnick, G., et al. 2006, ApJ, 650, 624
\reference{} Salpeter, E. E. 1955, ApJ, 121, 161
\reference{} Sanders, D. B.,\& Mirabel, I. F. 1996, ARA\&A, 34, 749
\reference{} Smail, I., Ivison, R. J.,\& Blain, A. W. 1997, ApJ, 490, L5
\reference{} Spinoglio, L., Malkan, M. A., Rush, B., Carrasco, L.,\& Recillas-Cruz, E. 1995, ApJ, 453, 616
\reference{} Strolger, L.-G., et al. 2004, ApJ, 613, 200
\reference{} STScI, 2001, VizieR Online Data Catalog, 1271
\reference{} Szokoly, G. P., et al. 2004, ApJS, 155, 271
\reference{} Vandame, B. 2002, SPIE, 4847, 123
\reference{} Vandame, B., et al. 2001, astro-ph/0102300
\reference{} van der Wel, A., Franx, M., van Dokkum, P. G., Rix, H.-W., Illingworth, G. D.,\& Rosati, P. 2005, ApJ, 631, 145
\reference{} van Dokkum, P. G., et al. 2006, ApJ, 638, 59
\reference{} Vanzella, E., et al. 2008, A\&A, 478, 83
\reference{} Williams, R. E., et al. 1996, AJ, 112, 1335
\reference{} Wolf, C., et al. 2004, A\&A, 421, 913
\reference{} Wuyts, S., et al. 2007, ApJ, 655, 51

}
\end {references}

\end {document}